\documentclass{article}

\usepackage{PRIMEarxiv}

\usepackage{algorithm}
\usepackage{algorithmic}
\usepackage{amsmath}
\usepackage{bm}
\usepackage{amsfonts}
\usepackage{makecell}
\usepackage{array}
\usepackage{amsmath}
\usepackage{booktabs}
\usepackage{longtable}
\usepackage{multirow}
\usepackage{booktabs}
\usepackage{tikz}
\usepackage{pgfplots}
\usepackage{xcolor}
\usepackage{subcaption}
\usepackage{underscore}
\usepackage{array}
\usepackage{tabularx}
\usepackage{booktabs}
\usepackage{multirow}
\usepackage{caption}  % in preamble for \captionof
\usepackage[flushleft]{threeparttable}
\usepackage[table,xcdraw]{xcolor}
\usepackage{pifont}
\newcommand{\cmark}{\ding{51}}
\newcommand{\xmark}{\ding{55}}
\usepackage{balance} % for balancing columns on the final page
\usepackage[utf8]{inputenc} % allow utf-8 input
\usepackage[T1]{fontenc}    % use 8-bit T1 fonts
\usepackage{hyperref}       % hyperlinks
\usepackage{url}            % simple URL typesetting
\usepackage{booktabs}       % professional-quality tables
\usepackage{amsfonts}       % blackboard math symbols
\usepackage{nicefrac}       % compact symbols for 1/2, etc.
\usepackage{microtype}      % microtypography
\usepackage{lipsum}
\usepackage{fancyhdr}       % header
\usepackage{graphicx}       % graphics
\usepackage{paralist}  
\usepackage{bm}  % For bold math symbols
\usepackage{placeins}

\usepackage{amsmath} 
\usepackage{multirow} 
\usepackage{tabularx}
\usepackage{booktabs}
\usepackage{float}
\usepackage{multirow}
\usepackage[table,xcdraw]{xcolor}

\usepackage{threeparttablex} % works with tabularx
\usepackage{pifont}

\newcolumntype{Y}{>{\centering\arraybackslash}X}
\usepackage[numbers]{natbib} % or [authoryear] if you prefer author-year

\graphicspath{{media/}}     % organize your images and other figures under media/ folder

\title{Skill-Aligned Fairness in Multi-Agent Learning for Collaboration in Healthcare
\thanks{\textit{\underline{Citation}}: 
\textbf{Promise Osaine Ekpo, Brian La, Thomas Wiener, Saesha Agarwal, Arshia Agrawal, Gonzalo Gonzalez-Pumariega, Lekan P.\ Molu, Angelique Taylor. Skill-Aligned Fairness in Multi-Agent Learning for Collaboration in Healthcare. Pages.... DOI:000000/11111.}} 
}

\author{
  Promise Ekpo \\
  Cornell Tech, NY, USA \\
  \texttt{poe6@cornell.edu} \\
  \And
  Brian La \\
  Cornell Tech, NY, USA \\
  \texttt{byl8@cornell.edu} \\
  \And
  Thomas Wiener \\
  Cornell Tech, NY, USA \\
  \texttt{tfw29@cornell.edu} \\
  \And
  Saesha Agarwal \\
  Cornell Tech, NY, USA \\
  \texttt{sa2388@cornell.edu} \\
  \And
  Arshia Agrawal \\
  Cornell Tech, NY, USA \\
  \texttt{aa2443@cornell.edu} \\
  \And
  Gonzalo Gonzalez-Pumariega \\
  Cornell Tech, NY, USA \\
  \texttt{gg387@cornell.edu} \\
  \And
  Lekan \ Molu \\
  Microsoft Research, NY, USA \\
  \texttt{patlekano@gmail.com} \\
    \And
  Angelique Taylor \\
  Cornell Tech, NY, USA \\
  \texttt{amt298@cornell.edu} \\
}

\begin{document}
\maketitle

\begin{figure*}[h]
  \centering
  \includegraphics[width=\textwidth]{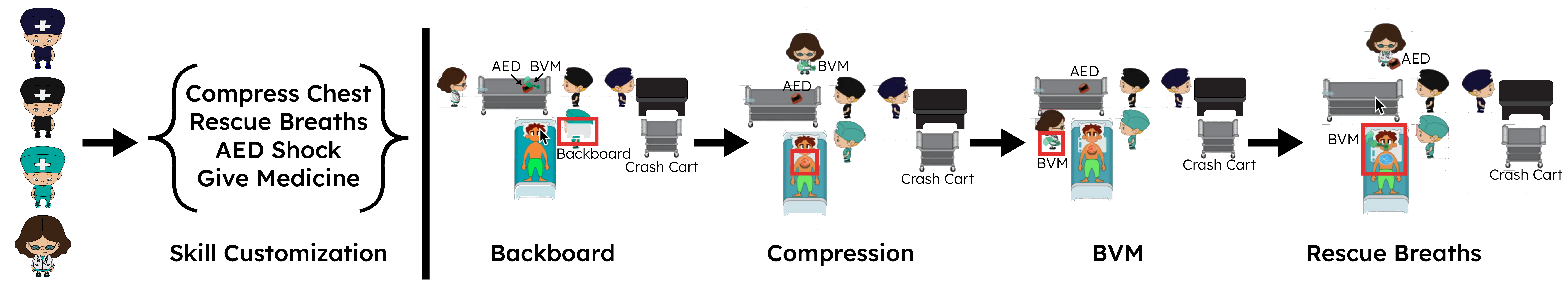}
  \caption{\textbf{MARLHospital Environment.} The environment integrates a PDDL planner with a MARL state layer to model skill-aligned fairness and shared-task coordination among healthcare workers. The goal is to pick the backboard from the crash cart, move to the patient, place it under the patient, compress patient chest multiple times, retrieve the Bag-Valve-Mask (BVM) from the crash cart, give patient rescue breaths multiple times.}
  \label{fig:teaser}
\end{figure*}

%   \includegraphics[width=\textwidth]{LaTeX/images/teaser_image_v2.png}
%   \caption{\textbf{MARLHospital Environment.} The environment integrates a PDDL planner with a MARL state layer to model skill-aligned fairness and shared-task coordination among healthcare workers. The goal is to pick the backboard from the crash cart, move to the patient, place it under the patient, compress patient chest multiple times, retrieve the Bag-Valve-Mask (BVM) from the crash cart, give patient rescue breaths multiple times.}
%   \Description
%   {\textbf{MARLHospital Overview.} The environment integrates a PDDL planner with a MARL state layer to model skill-aligned fairness and shared-task coordination among healthcare workers. The goal is to pick the backboard from the crash cart, move to the patient, place it under the patient, compress patient chest multiple times, retrieve the Bag-Valve-Mask (BVM) from the crash cart, give patient rescue breaths multiple times. Then, integrate a PDDL planner with a MARL state layer to model skill-aligned fairness and shared-task coordination among healthcare workers.}

\begin{abstract}
Fairness in Multi-Agent reinforcement learning (MARL) is often framed as a workload balance problem, overlooking agent expertise and the structured coordination required in real-world domains. 
In healthcare, equitable task allocation requires both workload balance and skill-task alignment to prevent burnout and overuse of highly skilled agents. 
We make two contributions to address this problem. 
First, we introduce MARLHospital, a customizable healthcare-inspired environment for training and evaluating team compositions and energy-constrained scheduling impacts on fairness using state-of-the-art MARL algorithms. 
Second, we propose FairSkillMARL, a framework that defines fairness as the dual objective of workload balance and skill-task alignment. 
%Across fairness weights, 
FairSkillMARL achieves higher alignment and comparable or lower workload disparity than workload-only baselines, improving coordination by up to 60\%  without reducing success rates. 
Results show that aligning effort with expertise enables more coordinated and fair multi-agent behavior. Our work provides tools and a foundation for studying fairness in heterogeneous multi-agent systems.
% Our work provides tools and a foundation for studying fairness in heterogeneous multi-agent systems where aligning effort with expertise is critical.
\end{abstract}

% keywords can be removed
\keywords{Multi-agent systems \and Healthcare \and Fairness \and Benchmark}

\section{Introduction}
\label{intro}
% \lipsum[2]
% \lipsum[3]

%%%%%%%%%%%%%%%%%%%%%%%%%%%%%%%%%%%%%%%%%%%%%%%%%%%%%%%%%%%%%%%%%%%%%%%%%%%%%%%%%%%%%%%%%%
%%%%%%%%%%%%%%% Big idea, set the stage. 1-2 paragraphs max

In multi-agent systems, agents must learn to cooperate while leveraging resources in their environment to achieve individual and collective objectives \cite{huh_multi-agent_2024}. In this paper, we are interested in effectively modeling coordination among healthcare workers (HCWs) as a multi-agent system in safety-critical environments, such as the Emergency Department (ED). A distinguishing feature of these environments that necessitates new tools to model behavior is the need for workers to perform shared tasks and occasionally relieve other workers based on availability, effort, and skills.

%In multi-agent systems, agents must learn to cooperate while leveraging resources in their environment to achieve individual and collective objectives \cite{huh_multi-agent_2024}. In this paper, we are interested in effective modeling methods in safety-critical healthcare workers (HCW) environments, such as emergency departments (ED). A distinguishing feature of these environments that necessitates new tools to model behavior is the need for workers (henceforth referred to as agents) to perform overlapping tasks, and occasionally relieve other workers based on availability, energy levels, and skills.
%\newline

%%%%%%%%%%%%%%%%%%%%%%%%%%%%%%%%%%%%%%%%%%%%%%%%%%%%%%%%%%%%%%%%%%%%%%%%%%%%%%%%%%%%%%%%%%
%%%%%%%%%%%%%%% OMG PROBLEM: Oh no! If we don’t solve this, yikes!
Common medical procedures in the ED are composed of high-level tasks and subtasks that must be carried out sequentially to achieve a goal. 
This includes \textit{individual tasks} such as the use of an automated external defibrillator and administering medication, and \textit{shared tasks} such as Cardiopulmonary resuscitation (CPR); CPR  requires multiple agents to perform task-switching once they become fatigued from performing chest compressions.
In real-world environments, many HCWs are overworked (i.e., burnout) when assigned tasks that do not align with their expertise (i.e., skill-task misalignment).
When this happens, HCWs spend more time accomplishing tasks compared to a skilled HCW with relevant expertise who can perform the task more efficiently \cite{taylor_rapidly_2025, taylor_towards_2024, taylor_coordinating_2019}.
To prevent burnout and to ensure fairness in EDs, in this work, we propose a framework for fairness in a multi-agent reinforcement learning (MARL) setting.

%Common medical procedures in the ED are composed of subtasks that must be carried out sequentially to achieve a goal. This  includes  \textit{individual tasks} such as the use of automated external defibrillator and administering medication, and \textit{shared tasks} such as Cardiopulmonary resuscitation (CPR); CPR  requires multiple agents to perform task-switching once they become fatigued from performing chest compressions. In real-world environments, many HCWs are overworked (i.e., burnout) when assigned tasks that do not align with their expertise (i.e., skill-task misalignment). When this happens, HCWs spend more time accomplishing tasks compared to a skilled HCW with relevant expertise who can perform the task more efficiently \cite{taylor_rapidly_2025, taylor_towards_2024, taylor_coordinating_2019}. To prevent burnout and to ensure fairness in EDs, in this work, we propose a framework for fairness in a multi-agent reinforcement learning (MARL) setting.

% \begin{figure}[t]
% \centering
%     \includegraphics[width=0.4\textwidth]{images/teaser_image.png}
%     \caption{\textbf{MARLHospital}: Breakdown of the Planning Domain Definition Language (PDDL) environment wrapped by the MARL environment with 3 HCWs and a patient.}
%     \label{fig:main_figure_draft}
% \end{figure}

%%%%%%%%%%%%%%%%%%%%%%%%%%%%%%%%%%%%%%%%%%%%%%%%%%%%%%%%%%%%%%%%%%%%%%%%%%%%%%%%%%%%%%%%%%
%%%%%%%%%%%%%%% BRIEFLY (1-2 paragraphs) what has been done before
%%%%%%%%%%%%%%% OMG GAP: Here is where you outline the missing part of what’s been done before. 

As a first step towards a fairness framework for MARL for varying-skilled heterogeneous agents, such as in the ED, we have three requirements.
First, we require a multi-agent simulation environment that captures the challenges faced by HCWs during medical procedures. Prior classes of MARL algorithms, including Independent learning (IL) \cite{tampuu_multiagent_2015,witt_is_2020} and Centralized Training with Decentralized Execution (CTDE) algorithms \cite{sunehag_value-decomposition_2017,rashid_qmix_2018}, have been benchmarked on cooperative applications where multiple agents must work together to achieve a goal; however, limited work addressed the unique challenges faced by healthcare workers in ED scenarios.

Second, simulators provide a controlled environment for modeling and analyzing cooperative tasks among multiple agents, thereby facilitating detailed optimization and evaluation of team dynamics and coordination \cite{carroll_utility_2020,wang_demo2code_2023,samvelyan_starcraft_2019,ellis_smacv2_2023}; yet, existing simulators lack realistic healthcare-themed environments that model agents' level of energy while taking actions, team compositions, and expertise.

Third, prior work has quantified fairness as reward equality in social dilemmas \cite{hughes_inequity_2018}, worst-case performance guarantees in traffic scheduling  \cite{yuan_multi-agent_2021}, balanced throughput in network control \cite{fang_fairness-aware_2024,huang_fairness-aware_2022}, equitable effort distribution in cooperative navigation tasks \cite{aloor_cooperation_2024}, cumulative reward parity in hierarchical learning (such as Fair-Efficient Networks, \cite{ jiang_learning_2019} and demographic parity in resource allocation \cite{malfa_fairness_2025}. 
However, these definitions do not jointly account for the skill-task alignment and overutilization of agents who are repeatedly tasked with the most demanding actions.
Furthermore, existing simulators lack realistic healthcare-themed environments that model agent energy levels, skill-task alignment, and workload in structured team configurations. 
More broadly, the intersection of MARL, healthcare, and fairness is understudied.
% In this paper, we address this gap by introducing MARLHospital, a simulation environment that captures HCWs' interactions during medical procedures. 
% MARLHospital is beneficial because it models structured teams with varying expertise, in tasks inspired by real-world medical procedures, and uniquely includes the execution of order-dependent and shared-task mode where agents alternate actions due to energy constraints, modeling agent fatigue across long time horizons, unlike prior work \cite{carroll_utility_2020,wang_demo2code_2023,samvelyan_starcraft_2019,ellis_smacv2_2023}.

\begin{table*}[t]
    \centering
    \caption{Comparison between \textit{MARLHospital} and other multi-agent simulators.}
    \label{tab:simulators_updated_fairppo}
    \begin{threeparttable}
        \small
        \setlength{\tabcolsep}{5pt}
        \begin{tabularx}{\textwidth}{@{}l*{6}{>{\centering\arraybackslash}X}@{}}
            \toprule
            & MARL & Partial-Obs. & Hospital Setting & Skill Specialization & Variable Skills & Shared Tasks \\ 
            \midrule
            Overcooked-AI \cite{carroll_utility_2020} & \xmark & \xmark & \xmark & \xmark & \xmark & \xmark \\
            Agent Hospital \cite{li_agent_2024} & \xmark & \cmark & \cmark & \xmark\tnote{1} & \xmark & \xmark \\
            MARL Robotic Surgery \cite{scheikl_cooperative_2021} & \cmark & \cmark & \cmark\tnote{2} & \cmark & \xmark & \xmark \\
            MARL for Nurse Rostering \cite{zhang_multi-agent_2024} & \cmark & \xmark & \cmark\tnote{3} & \xmark & \xmark & \xmark \\
            Cuisine World \cite{gong_mindagent_2023} & \xmark & \xmark & \xmark & \xmark & \xmark & \xmark \\
            VirtualHome \cite{puig_virtualhome_2018} & \xmark & \xmark & \xmark & \xmark & \xmark & \xmark \\
            SMACv2 \cite{ellis_smacv2_2023} & \cmark & \cmark & \xmark & \cmark & \xmark & \xmark \\
            Pommerman \cite{resnick_pommerman_2018} & \cmark & \cmark & \xmark & \xmark & \xmark & \xmark \\
            Melting Pot 2.0 \cite{agapiou_melting_2023} & \cmark & \cmark & \xmark & \cmark & \xmark & \xmark \\
            Robotouille \cite{wang_demo2code_2023} & \xmark & \xmark & \xmark & \xmark  & \xmark & \xmark \\
            HMARL for Medical Allocation \cite{hao_hierarchical_2023} & \cmark & \xmark\tnote{4} & \xmark & \xmark & \xmark & \xmark \\
            MA Hospital Infection Sim \cite{esposito_multi-agent_2020} & \xmark & \xmark & \cmark & \cmark  & \xmark & \xmark \\
            H2-MARL \cite{luo_h2-marl_2025} & \cmark & \cmark & \cmark & \xmark & \xmark & \xmark \\
            NurseSchedRL \cite{koduri_nurseschedrl_2025} & \cmark & \xmark & \cmark & \cmark\tnote{7} & \xmark & \xmark \\
            MedScrubCrew \cite{ruiz_mejia_medscrubcrew_2025} & \xmark & \xmark & \cmark & \xmark & \xmark & \xmark \\
            Multi-Agent Collaborative Work Sim \cite{torres-huesca_multi-agent_2021} & \xmark & \xmark & \cmark & \cmark & \xmark & \xmark \\
            Hierarchical MA Resource Allocation \cite{hao_hierarchical_2023} & \cmark & \xmark\tnote{4} & \cmark & \xmark & \xmark & \xmark \\
            Fair-PPO HospitalSim \cite{malfa_fairness_2025} & \cmark & \xmark\tnote{5} & \cmark & \xmark\tnote{6} & \xmark & \xmark \\
            \midrule
            \textit{MARLHospital (Ours)} & \cmark & \cmark & \cmark & \cmark & \cmark & \cmark \\
            \bottomrule
        \end{tabularx}
        \begin{tablenotes}
            \item[1] Patients, doctors, and nurses modeled; no specialization among HCWs (doctors are the only “trained” agents).
            \item[2] Simulates surgical actions in an OR; does not represent full hospital topology.
            \item[3] Focused on combinatorial scheduling; no patient treatment interactions.
            \item[4] Imperfect information handled via custom uncertainty; not formulated as an explicit POMDP.
            \item[5] Event-driven flow-based simulation of daily hospital logistics; no timestep-based POMDP structure.
            \item[6] Fairness across patient groups, not worker skill heterogeneity.
            \item[7] Uses fixed nurse skill profiles for assignment; profiles do not evolve during an episode.
        \end{tablenotes}
    \end{threeparttable}
\end{table*}

To address these research gaps, we introduce MARLHospital, a simulation environment that captures HCWs' interactions during medical procedures in Figure \ref{fig:teaser}. %\cite{taylor_hospitals_2022}.
MARLHospital is beneficial because it models structured teams with varying expertise, in tasks inspired by real-world medical procedures, and uniquely includes the execution of order-dependent and shared-task mode where agents alternate actions due to energy constraints, modelling agent fatigue across long time horizons, unlike prior work \cite{carroll_utility_2020,wang_demo2code_2023,samvelyan_starcraft_2019,ellis_smacv2_2023}. 
Furthermore, we introduce FairSkillMARL, a framework that formulates fairness as a balance of workload and skill-task alignment during collaborative tasks, whilst factoring in the progressively dissipative energy levels of agents. 
This framing has the potential to improve the efficiency of healthcare teams by leveraging task-switching and agent expertise.

Our \textbf{contributions} are threefold: 
\begin{inparaenum}[1)]
\item We introduce a customizable hospital-themed game environment inspired by real-world settings. Unlike most existing MARL simulators, which target abstract coordination or traffic control, our environment models sequential, order-dependent medical procedures and structured team roles, filling a critical gap for evaluating fairness in healthcare applications.
\item  We share insights about the performance of four standard MARL algorithms, including on-policy and off-policy methods, across varying team compositions, task difficulties, and energy levels in three healthcare benchmark tasks. 
\item  We introduce a FairSkillMARL framework, which redefines fairness as a composite disparity of agent skill-task alignment and workload, and our results demonstrate that FairSkillMARL shows competitive performance alongside state-of-the-art fairness metrics, particularly for simulated tasks and the need for more robust metrics to capture skill misalignment. We extend fairness in MARL beyond workload balance to include skill-task alignment 
% and energy constraints,
introducing new metrics and a healthcare-inspired benchmark that enables the evaluation of cooperative policies in realistic, heterogeneous team settings. 
\end{inparaenum}

\section{Related Work}
\label{library}

\textbf{Multi-agent systems.} Multi-agent systems model interactions between multiple agents in cooperative, competitive, and mixed settings. Our research explores two classes of multi-agent reinforcement learning (MARL) algorithms, Independent Q-Learning (IQL) \cite{tampuu_multiagent_2015} and Centralized Training with Decentralized Execution (CTDE) \cite{amato_first_2024}.

IQL agents simultaneously learn a Q-function within the same environment based solely on its local observations and actions, and \textit{Independent Proximal Policy Optimisation} (IPPO) \cite{witt_is_2020}, which applies the popular Actor-Critic method PPO \cite{schulman_proximal_2017} independently to each agent using its own observations and actions. While these methods provide a straightforward and effective baseline in practice, the lack of global state information during training may lead to difficulty in learning coordination as unstable learning may stem from the problem of environment non-stationarity.

In CTDE \cite{amato_first_2024}, agents are trained using shared global information but learn policies conditioned only on their local observations, enabling effective cooperative behavior while maintaining individual decision-making during execution time.
While some CTDE algorithms focus on factorizing the team's joint action-value function into multiple individual action-value functions that can be used during deployment such as \textit{Value Decomposition Networks} (\textit{VDN}) \cite{sunehag_value-decomposition_2017}, \textit{QMIX} \cite{rashid_qmix_2018}, both Q-based learning methods that backpropagate global gradients to local agent networks. 
Another category of CTDE algorithms uses centralized policy gradient methods involving individual actors and centralized critics (e.g \textit{Multi-Agent PPO} (\textit{MAPPO}) \cite{yu_surprising_2022} and \textit{Counterfactual Multi-Agent Policy Gradients (COMA)} \cite{foerster_counterfactual_2018}).
More relevant to our setting, recent MARL applications in healthcare have used Hierarchical MARL to delegate task specialization effectively, \cite{hao_hierarchical_2023, perera_demystifying_2023, tan_advancing_2024}, but this framework assumes centralized decision making will be possible at deployment.

\textbf{Multi-agent environments.} To understand interactions between multiple agents, the majority of research efforts have captured popular settings such as games \cite{guss_minerl_2019,carroll_utility_2020} and cooking environments \cite{gonzalez-pumariega_robotouille_2025}. 
Despite their successes, many approaches did not take into account the need for fully customizable skill levels across agents or the evaluation of predefined team compositions inspired by human-human collaboration in safety-critical environments. 
Although simulators such as Overcooked-AI \cite{carroll_utility_2020} and Robotouille \cite{gonzalez-pumariega_robotouille_2025} model tasks with temporal dependencies, they assume symmetric agent expertise and lack support for evaluating different team compositions. 
SMACv2 \cite{ellis_smacv2_2023} supports heterogeneous agents but does not support structured task hierarchies and the shared task mode in MARLHospital. 
While Pommerman \cite{resnick_pommerman_2018}, CUISINEWORLD \cite{gong_mindagent_2023}, and Melting Pot \cite{agapiou_melting_2023} focus on abstract coordination, they do not capture team heterogeneity or domain-specific expertise. 
VirtualHome \cite{puig_virtualhome_2018} supports long-horizon tasks but lacks real-time multi-agent coordination, and AgentHospital \cite{li_agent_2024} focuses on dialogue using Large Language Model rather than MARL benchmarking. We make a detailed comparison between MARLHospital and others in Table \ref{tab:simulators_updated_fairppo}.

\textbf{Fairness in multi-agent systems.} Prior work has quantified fairness across multi-agent domains. 
In social dilemmas, inequity aversion reduces envy and guilt by penalizing outcome disparities, promoting cooperation \citep{hughes_inequity_2018}. 
In networked systems, fairness is often framed through worst-case guarantees, namely, maximizing the 5\%-tile user data rate \citep{yuan_multi-agent_2021}. 
In traffic signal control, fairness is integrated into value functions to ensure low-traffic lanes are scheduled equally \citep{fang_fairness-aware_2024, huang_fairness-aware_2022}. 
These approaches focus on reward parity or throughput smoothing but assume homogeneous agents. Structural fairness has been explored in cooperative control by penalizing travel deviations to balance workload \citet{aloor_cooperation_2024} and propose switching between fair and efficient subpolicies \citet{jiang_learning_2019}. 
Social fairness constraints have also been encoded, such as demographic parity in Proximal Policy Optimization (PPO) \citep{malfa_fairness_2025}. 
While addressing various fairness goals, these works often overlook skill-agent compatibility and task-agent mismatch. 
In healthcare, fairness extends beyond efficiency to patient safety, requiring mechanisms that prevent HCWs fatigue from unfair workload and treatment delays caused by assigning tasks to workers outside their specialties.\newline
Recent studies in multi-agent systems have explored many ways to define and measure fairness, such as \textit{fair division}, \textit{fair allocation}, and \textit{fairness guarantees} in cooperative and learning settings. Research on \textit{envy-freeness} has examined how to divide limited resources so that no agent prefers another agent’s share~\cite{bredereck_computing_2025, barman_parameterized_2024}. Other work has focused on \textit{resilient} and \textit{group-based} fairness rules that ensure stable and inclusive outcomes~\cite{mutzari_resilient_2023, scarlett_for_2023}. Fairness has also been studied in learning environments, such as fairness-aware reinforcement learning~\cite{smit_fairness_2024} and fairness-of-exposure in decision-making systems~\cite{sood_fairness_2024}. More recently, FAPPO~\cite{siddique_towards_2025} explored how to balance fairness and efficiency in cooperative MARL. Our \textbf{FairSkillMARL} framework builds on these directions by focusing on \textit{skill–task alignment}, which captures how heterogeneous agents can share tasks fairly while balancing their workloads.

\section{MARLHospital}
\label{marl}

We present MARLHospital, a framework for modeling multi-agent collaborative tasks in medical environments. 
The MARLHospital environment builds upon Robotouille \cite{gonzalez-pumariega_robotouille_2025} using the EPyMARL  \cite{papoudakis_benchmarking_2021} setup for MARL algorithms. 
Unlike prior work, MARLHospital is a customizable multi-agent environment in terms of team compositions, shared vs. individually task modes.
We draw inspiration from real-world safety-critical environments (i.e., the ED) in the design of MARLHospital where HCWs have varying skills, energy levels and workloads.

%We present MARLHospital, a framework for modeling multi-agent collaborative tasks in medical environments. %We present our framework for modeling multi-agent collaborative tasks in medical environments using the proposed MARLHospital simulator. 
%We address the challenge of coordinating agents with varying expertise levels and in shared task mode, which is prevalent in multi-agent settings, particularly in ERs.

\subsection{Notations and Preliminaries}
\phantomsection
\label{sec:problem-formulation}

We formulate MARLHospital as a decentralized partially observable Markov decision process (Dec-POMDP)~\cite{oliehoek_concise_2016}, represented by the tuple $(I,\mathcal{S},\mathcal{A},T,R,\Omega,O,\gamma)$. 
Here $I=\{1,\dots,n\}$ denotes the set of agents; $\mathcal{S}$ is the global state space; $\mathcal{A}$ and $\Omega$ are the joint action and observation spaces; $T$ is the transition kernel; $R$ the team reward; $O$ the observation kernel; and $\gamma\in[0,1)$ the discount factor. 
$\mathcal{A}$ and $\Omega$ are the joint action and observation spaces because the transition and observation depend on all agents acting simultaneously, the environment evolves based on their combined actions and produces observations for each of them.

Each agent $i\in I$ has a discrete local action space $\mathcal{A}_i$ e.g. where $a_i$ is drawn from $\mathcal{A}_i = \{ 1,2,3\} $ and an observation space $\Omega_i$.  
%$\mathcal{A}_i$
At each timestep $t$, the environment is in state $\mathbf{s}_t\in\mathcal{S}$, each agent observes $\mathbf{o}^i_t\in\Omega_i$, and selects an action $\mathbf{a}^i_t\in\mathcal{A}_i$ according to its policy $\pi^i$. 
The team jointly executes $\mathbf{a}_t=(a_t^1,\ldots,a_t^n)$ and receives the next state $\mathbf{s}_{t+1}$ drawn from $T(\mathbf{s}_{t+1}\mid\mathbf{s}_t,\mathbf{a}_t)$ and a shared reward $R(\mathbf{s}_t,\mathbf{a}_t)$. 
Episodes terminate when all medical subtasks are complete or a time limit is reached.

The global state $\mathbf{s}_t$ encodes all medically relevant information in a resuscitation scenario, combining symbolic, spatial, and continuous The global state includes the patient’s current state (whether chest-compressed or rescue-breathed), the locations of the three agents across six discrete stations (\texttt{hospital\_cart\_right},\texttt{hospital\_cart\_left},\texttt{table},\texttt{small_hospital\_cart}, \texttt{patient\_legs}, \texttt{patient\_bed\_station}), each agent’s held item (\texttt{BVM}, \texttt{Backboard}, or \texttt{patient}), their discrete skill levels for chest compression and giving rescue breathes, continuous energy levels $e_{i,t}\in[0,e_{\max}]$ tracking fatigue (The subscript 
i,t denotes both the agent index and the timestep, which is standard in Dec-POMDP notation. It helps distinguish each agent’s energy value as it changes over time.), and binary indicators of whether CPR or rescue-breath goals have been reached. 
The complete state vector has 174 binary features and three continuous components. 
Each agent observes only a local subset $\mathbf{o}^i_t$ containing its own position, held item, skill indicators, energy level, and partial patient information; other agents’ states remain hidden, enforcing partial observability.

Each agent’s action space $\mathcal{A}_i$ consists of eight discrete primitives: 
\texttt{move}, \texttt{pick}, \texttt{place}, \texttt{stack}, \texttt{treat(patient)}, \texttt{compress\_chest},\newline \texttt{give\_rescue\_breaths}, and \texttt{noop}. 
Actions are symbolic commands grounded in the PDDL simulator. 
The \texttt{compress\_chest} action is an energy-constrained shared task controlled by the fatigue model in Section~\ref{sec:exp-rss}, while the remaining actions alter symbolic or spatial states without energy cost. 
The joint action $\mathbf{a}_t$ therefore represents the simultaneous choices of all $n$ agents, with $|\mathcal{A}_i|=8$ and $|\mathcal{A}|=8^n$ for the three-agent setting used in our experiments.

Rewards are defined through a heuristic progress function $H(\mathbf{s}_t)$ that measures advancement toward subgoals such as positioning the CPR board, picking the right items, placing the item at the right station, and completing chest compressions or administering rescue breaths. 
The immediate reward at timestep $t$ is
\[
R(\mathbf{s}_t,\mathbf{a}_t) = H(\mathbf{s}_t) - H(\mathbf{s}_{t-1}),
\]
yielding positive values only when the team makes progress toward medical completion. 
Each policy $\pi^i$ seeks to maximize the expected discounted return $G_t=\sum_{k=t}^T \gamma^{k-t}R(\mathbf{s}_k,\mathbf{a}_k)$.

%The MARLHospital environment builds upon Robotouille \cite{gonzalez-pumariega_robotouille_2025} using the EPyMARL  \cite{papoudakis_benchmarking_2021} setup for MARL algorithms. 

\subsection{Team Compositions} 

Understanding team composition is essential for modeling clinical settings, as the structure and capabilities of a care team influence task performance, decision speed, and workload distribution. 
We define three team compositions that vary in expertise levels (EL) of agents: \textit{uniform}, \textit{specialized}, and \textit{interdependent} teams. 
In \textit{uniform Teams}, all agents possess identical capabilities and can perform every subtask with equal efficiency. 
This structure allows maximum flexibility as any agent can complete any part of the task independently, and coordination is optional.
\textit{Specialized Teams} allows agents to be more efficient in one particular subtask but still retains the ability to perform all others, although more slowly. 
This specific setting incentivizes but does not necessitate collaboration.
In \textit{Interdependent Teams (Forced Cooperation)}, agents are capable of performing only a subset of subtasks and cannot execute at least two other tasks. 

\begin{figure}[t]
\centering
    \includegraphics[width=0.52\textwidth]{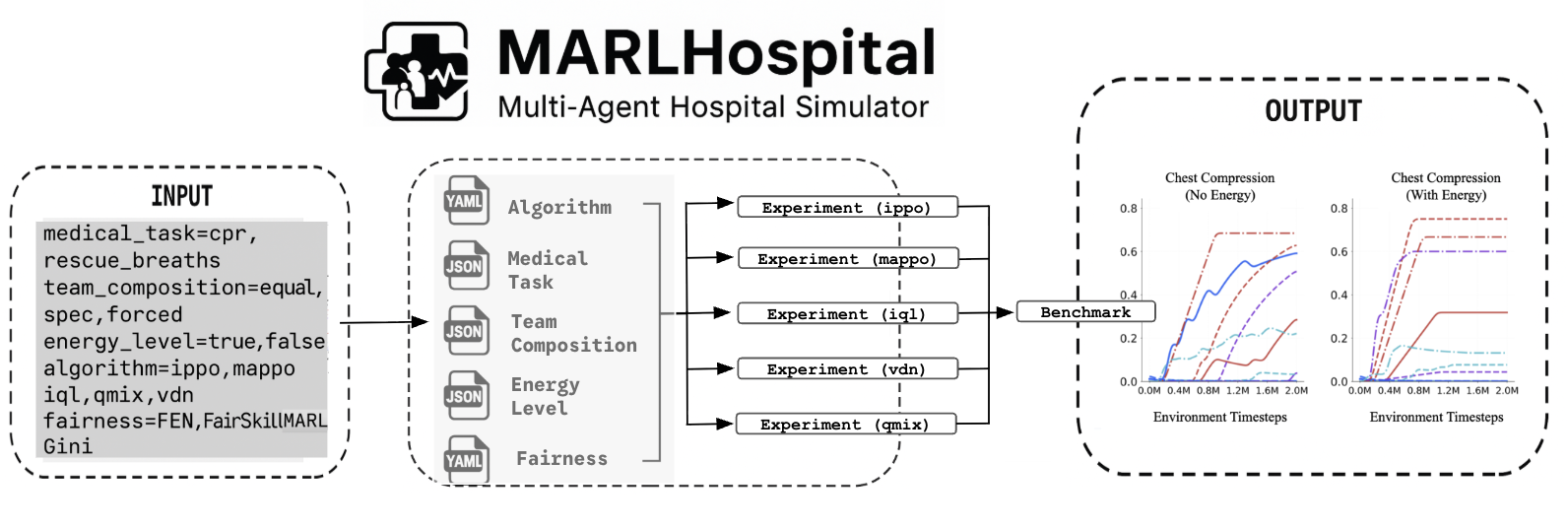}
    \caption{MARLHospital  Execution flow.}
    \label{fig:task_execution}
\end{figure}

\subsection{Medical Tasks} 
\label{medical-tasks}

We model the CPR and AED tasks using the standard procedural steps of ``Adult Basic Life Support'' from the American Red Cross code cards in Figure \ref{fig:task_difficulty_v3} with the full execution flow in Figure \ref{fig:task_execution}. 
These cards provide visual flow charts guiding resuscitation procedures to capture real-world tasks based on evidence-based practices \cite{noauthor_american_nodate}. 
These procedural steps are modeled in the action space of MARLHospital.

\subsubsection{Agent Energy Levels for Shared Tasks}
\label{sec:exp-rss}
\begin{figure}[t]
\centering
    \includegraphics[width=0.52\textwidth]{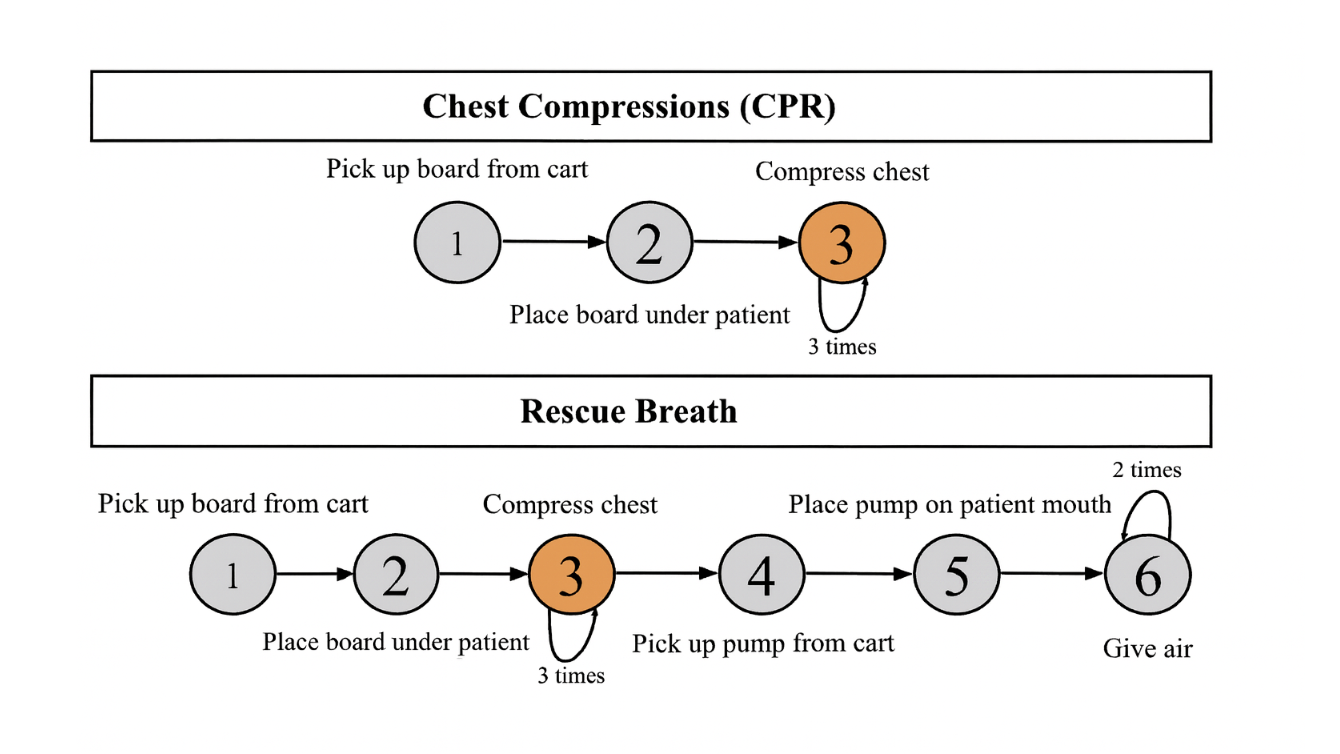}
    \caption{Task flow diagrams for the check compression (CPR) and rescue breath tasks in MARLHospital. The yellow-shaded subtask supports shared action with energy constraints.}
    \label{fig:task_difficulty_v3}
\end{figure}

Cardiopulmonary resuscitation (CPR) in real-world practice requires healthcare workers (HCWs) to alternate every two minutes to prevent fatigue \cite{savatmongkorngul_comparison_nodate}. To reflect this, MARLHospital introduces an \textbf{energy model} that constrains how long agents can continuously perform effort-intensive \emph{\textbf{shared tasks}}, such as chest compressions (Figure~\ref{fig:task_difficulty_v3}). This mechanism enforces natural alternation between agents based on their remaining energy levels. Each agent $i\in I$ maintains an energy variable $e_{i,t}$ at timestep $t$, where $0\le e_{i,t}\le e_{\max}$ and $e_{\max}$ is the maximum possible energy. The initial energy levels of all agents, as well as the number of chest compressions required to reach the goal, are configurable in the simulator. Performing an action $a\in\mathcal{A}_i$ incurs an energy cost $c_a$: it is positive ($c_a\ge0$) for energy-consuming actions that must be executed at maximum efficiency (e.g., \texttt{compress\_chest}) and negative ($c_a<0$) for restorative or idle actions (e.g., a ``no-op" that allows recovery or any non-yellow action in Figure~\ref{fig:task_difficulty_v3}). The energy dynamics for each agent follow
\begin{equation}
e_{i,t+1}=\min\left(e_{\max},\,e_{i,t}-c_a\right),
\label{equ:energy_update}
\end{equation}
where $c_a$ takes positive values for costly actions and negative values for recovery. When the simulator is in \emph{shared-task mode}, where the energy constraint applies to the \texttt{compress\_chest} action, an agent can only execute that action if it has sufficient remaining energy:
\begin{equation}
a_t^i=
\begin{cases}
\texttt{compress\_chest}, & \text{if } e_{i,t}\ge c_{\text{compress}},\\[3pt]
\texttt{noop}, & \text{otherwise.}
\end{cases}
\label{equ:energy_constraint}
\end{equation}
When the energy cost $c_{\text{compress}}$ is large relative to the recharge rate $|c_{\text{noop}}|$, agents must alternate more frequently to sustain task completion. This energy model is integrated into the Dec-POMDP formulation through the state vector $\mathbf{s}_t$, which includes each agent’s current energy $e_{i,t}$. By modeling fatigue directly within the state space, MARLHospital enables policies that learn both efficient task scheduling and energy-aware coordination over time.

%\subsection{Medical Tasks} 
%\label{medical-tasks}

%We model the CPR and AED tasks using the standard procedural steps of ``Adult Basic Life Support'' from the American Red Cross code cards (see Figure \ref{fig:task_difficulty_v3}). 
%These cards provide visual flow charts guiding resuscitation procedures to capture real-world tasks based on evidence-based practices \cite{noauthor_american_nodate}. 
%These procedural steps are modeled in the action space of MARLHospital. 

\subsubsection{Task Difficulty} We define the two goals as Partial (P) and Complete (C) with task difficulties based on the length of the time horizon. 
For the CPR goal, the HCWs must perform CPR, a short-time-horizon task that consists of picking up and placing a board under the patient and giving N chest compressions. 
For the rescue breaths, which is a longer horizon task, in addition to the CPR goal, the HCWs must perform the same actions in the CPR task, in addition to picking up the BVM and placing it on the patient to give them oxygen. 
Since we use procedural generation in the environment implementation, these goals can be modified in the environment configuration file.

\section{FairSkillMARL}
\label{sec:fairmarl}

We introduce FairSkillMARL, a novel reward function that encourages agents to strike a balance between workload and skill-task alignment.
Unlike prior fairness definitions that primarily focus on the distribution of agent workloads \citet{aloor_cooperation_2024} and throughput parity \citep{malfa_fairness_2025}), in FairSkillMARL, agents are penalized for the disparity between agent workloads and the alignment between an agents' task and expertise.
A skilled agent completes complex tasks quickly with low error probability, while an unskilled agent requires more time and makes more mistakes on the same task.
Our reward formulation prevents skilled agents from being overloaded with all complex tasks (causing fatigue) while also preventing unskilled agents from becoming bottlenecks on tasks beyond their capabilities. 
This captures the essential coordination challenge in ED, where performance depends on matching tasks to appropriate skill levels.

%We modify the reward function from the preliminaries section to penalise both disparity in workload balance and skill-task misalignment in individual and shared tasks. Unlike prior definitions that primarily models fairness as the distribution of agent workloads \citet{aloor_cooperation_2024}, throughput parity \citep{malfa_fairness_2025}), we model the reality that task completion time and error rates depend on agent skill levels. A skilled agent completes complex tasks quickly with low error probability, while an unskilled agent requires more time and makes more mistakes on the same task. Our reward formulation prevents skilled agents from being overloaded with all complex tasks (causing fatigue) while also preventing unskilled agents from becoming bottlenecks on tasks beyond their capabilities. This captures the essential coordination challenge in ED, where performance depends on matching tasks to appropriate skill levels.

\subsection*{FairSkillMARL Objective Definition}
\label{sec:fairmarl-objective}

Building on the notation introduced in Section~\ref{sec:problem-formulation},
we now formalize the fairness objective that operates over the same agent and action spaces defined in the Dec-POMDP.
We define \textbf{FairSkillMARL} as a fairness-aware objective that jointly accounts for 
(1) workload balance across agents and (2) alignment between an agent's skills and the tasks it performs.
Let $I=\{1,\dots,n\}$ denote the set of agents and $\mathcal{T}$ the set of all subtasks completed in an episode.
Each subtask $t\in\mathcal{T}$ is executed by one agent $i_t\in I$.
Unless otherwise specified, indices $i,j\in I$ denote \emph{agents}, while $i_t$ refers specifically to the agent that executed subtask $t$.
\paragraph{\textbf{Workload imbalance }($L_1$).}
We measure workload disparity among agents using the Gini index, which quantifies the deviation in the number of subtasks handled by each agent.
Let $x_i = |\tau_i|$ denote the number of subtasks performed by agent $i$, 
and $\bar{x} = \frac{1}{n}\sum_{i=1}^{n} x_i$ the mean workload across all $n$ agents.
Here $i$ and $j$ are agent indices used for pairwise comparisons only and can not be equal.
The workload imbalance is defined as:
\begin{equation}
L_1 = 
\frac{\sum\limits_{i=1}^{n}\sum\limits_{j=1}^{n} |x_i - x_j|}
     {2n^2 \bar{x}}.
\label{equ:gini}
\end{equation}
Equivalently, omitting self-pairs,
\(
L_1 = \frac{\sum_{1\le i<j\le n}|x_i - x_j|}{n(n-1)\bar{x}}.
\)
$L_1 \in [0,1]$ where $L_1 = 0$ represents perfectly equal workload distribution,
and higher values indicate increasing inequality in the number of tasks completed by each agent.

\paragraph{\textbf{Skill--Task Misalignment ($L_2$).}}
We define $L_2$ as a measure of how far the team’s actual task assignments deviate 
from the optimal skill-to-task allocation achievable under perfect specialization.
Let $\mathcal{T}$ denote the set of all subtasks completed during an episode, 
and $I = \{1,2,\dots,n\}$ the set of agents.
Each subtask $t \in \mathcal{T}$ is executed by one agent $i_t \in I$, 
where $i_t$ indexes the \emph{executor} of that subtask.
For comparison, we let $j \in I$ index all \emph{candidate agents} who could have performed the same subtask.
We introduce a skill function $\mathcal{S}_j(t) \in [0,1]$ 
that represents the normalized proficiency or capability of agent $j$ on task $t$,
where higher values indicate greater expertise.
The best achievable proficiency for task $t$ across the team is therefore 
$\max_{j \in I} \mathcal{S}_j(t)$.
Formally, we define:
\begin{equation}
L_2 = 1 -
\frac{\sum\limits_{t \in \mathcal{T}} \mathcal{S}_{i_t}(t)}
     {\sum\limits_{t \in \mathcal{T}} \max\limits_{j \in I} \mathcal{S}_j(t)}.
\label{equ:l2}
\end{equation}
Here, $i_t$ and $j$ serve distinct roles: 
$i_t$ refers to the agent that actually executed task $t$, 
while $j$ ranges over all agents to determine the ideal benchmark skill for that task. 
The numerator measures the cumulative skill of the agents who performed each task,
while the denominator measures the best possible cumulative skill if each task were assigned 
to its most capable agent. 
A lower $L_2$ value indicates better skill--task alignment, 
with $L_2 = 0$ corresponding to perfect alignment (every task performed by the most skilled agent) 
and $L_2 = 1$ indicating complete misalignment (no tasks aligned with agent expertise).

\paragraph{\textbf{Composite disparity ($L_3$)}.}
To capture both workload balance and skill alignment, 
we define a composite fairness objective as a convex combination:
\begin{equation}
L_3 = \alpha L_1 + (1-\alpha)L_2,
\label{equ:l3}
\end{equation}
where $\alpha \in [0,1]$ controls the trade-off between workload balance ($L_1$) and skill--task alignment ($L_2$).
A higher $\alpha$ emphasizes equal task sharing, while a lower $\alpha$ prioritizes specialization consistent with agents’ expertise.

\paragraph{\textbf{Fairness-shaped reward.}}
Let $R:\mathcal{S}\times\mathcal{A}\rightarrow\mathbb{R}$ denote the base team reward function,
and let $\lambda>0$ control the strength of fairness regularization.
The overall fairness-shaped reward used during training is:
\begin{equation}
r_t = R(\mathbf{s}_t,\mathbf{a}_t) - \lambda L_3,
\label{equ:rt}
\end{equation}
where $\mathbf{s}_t\in\mathcal{S}$ and $\mathbf{a}_t\in\mathcal{A}$ represent the joint state and joint action at timestep $t$.
For decentralized methods, a per-agent shaped signal $r_t^i = R_i(\mathbf{s}_t) - \lambda L_3$ can also be used,
where $R(\mathbf{s}_t,\mathbf{a}_t)=\sum_i R_i(\mathbf{s}_t)$.
This formulation allows FairSkillMARL to flexibly integrate fairness regularization into both
Independent Learning (IL) and Centralized Training with Decentralized Execution (CTDE) algorithms,
encouraging policies that are both efficient and skill-aware.

\section{Experiment 1: MARLHospital}
\label{experiment1}

Our experiments were designed to answer the following questions: 
(1) How do varying task difficulty and different team compositions affect the performance of existing MARL algorithms and team efficiency?
(2) How well do standard MARL algorithms perform in the shared task mode with varying energy levels, and how does this performance vary across team compositions?

\subsection{Baseline Methods}
\label{scalability}

Our experiments include standard MARL algorithms, including IL and CTDE, trainable on our custom simulation environment as building blocks for more sophisticated algorithms (i.e., {EPyMARL library} \citep{papoudakis_benchmarking_2021}\footnote{https://github.com/uoe-agents/epymarl} and {PyMARL}
\citep{samvelyan_starcraft_2019}\footnote{https://github.com/oxwhirl/pymarl})  \cite{foerster_learning_2016,he_opponent_2016,du_liir_2019,amato_first_2024}.
We investigate structured team compositions, including standard baselines that might facilitate different coordination strategies based on their decentralized and centralized training setups.

We summarize the IL and CTDE algorithms below. 
\begin{itemize}
\item \textbf{IQL} \cite{tampuu_multiagent_2015} enables each IL agent to simultaneously learn its Q-function within the environment based on its local observation history and actions. 
% \item \textbf{IPPO} \cite{witt_is_2020} applies the Actor-Critic method PPO \cite{schulman_proximal_2017} independently to IL agents  observations and actions. As PPO's clipped surrogate objective limits policy update sizes, it can reuse trajectories in multiple epochs to balance exploration and stability.
\item \textbf{MAPPO} \cite{yu_surprising_2022} is a CTDE policy gradient algorithm. Based on the PPO framework, thus enabling multiple updates on the same training batch to improve sample efficiency and stability, but it also uses a centralised state-value critic function conditioned on the state of the global environment.
\item \textbf{VDN} \cite{sunehag_value-decomposition_2017} decomposes the CTDE team's joint Q-value function into a sum of individual Q-value functions. The joint Q-value is trained using the Deep Q-Network (DQN) algorithm \cite{mnih_playing_2013}, and thus gradients are backpropagated to the individual agent networks. 
\item \textbf{QMIX} \cite{rashid_qmix_2018} extends CTDE VDN by using a mixing network to combine individual agent Q-values non-linearly, allowing for more complex value function factorization. The mixing network is constrained to maintain monotonicity in the relationship between the agent-specific and global Q-values, ensuring the optimal local actions and corresponding global joint actions are the same. 
\end{itemize}

\subsection{Training and Testing Procedure}
\label{baselines}

We benchmark state-of-the-art IL and CTDE MARL algorithms in MARLHospital with three agents.
We performed task-specific hyperparameter tuning across four algorithms (see Supplemental Materials). 
We trained the baseline algorithms with four seeds for 50 timesteps per episode using off-policy algorithms for 2M timesteps and on-policy algorithms for 20M timesteps, similar to EPyMARL \cite{papoudakis_benchmarking_2021}.
We trained off-policy algorithms using an experience replay buffer for stabilized policy learning on one NVIDIA GPU with x86 CPUs when available; otherwise, we used 4 CPU cores on a compute system with two Intel Xeon Gold 6448Y processors (each with 32 cores, 2.10--4.10GHz, 60MB cache, PCIe 5.0), providing a total of 64 CPU cores.

\subsection{Measures}
\label{sec:marl_measures}

We use \textbf{success rate} as a performance metric to assess how often agents reach the goal. 
This metric allows us to understand the stability and final performance of the trained policy across all evaluation episodes. 
Concretely, we report the total number of goals reached out of all test episodes during test time over 4 seeds, with each evaluation consisting of 100 episodes. 

\subsection{Experimental Setup} 
\label{sec-exp1}

We explore the impact of task difficulty, team compositions, and agent energy levels on the success rate of agents as follows:
\begin{itemize}
\item \textbf{Task difficulty}: To understand the impact of task difficulty, the agents perform CPR, a short time horizon task and rescue breaths, a long time horizon task.
\item \textbf{Team compositions}: To assess the effect of different team compositions, the agents skills are configured as uniform, specialized, and forced cooperation.
\item \textbf{Agent energy levels}: To explore the benefits of agent energy levels in the MARLHospital environment, we explore the performance of agents with and without energy function to model routine task-switching during CPR in the shared task mode.
\end{itemize}

\subsection{MARLHospital Results}
\label{sec:Algorithmic_Performance}

\subsubsection{Impact of Task Difficulty on Performance.} 
The results in Table \ref{tab:combined_results} with test success rates show that the CTDE algorithm (VDN) outperforms the DTDE algorithms in the CPR goal by a significant gap as seen in Figure \ref{fig:all_scenarios_horizontal_comparison}.
Additionally, in the  rescue breaths goal, the CTDE (VDN, QMIX) algorithms outperform the DTDE algorithms. 
This indicates that CTDE approaches are more robust under increased task complexity. Also, 
we performed pairwise Welch’s \textit{t}-tests on success rates for the chest compression task. \textbf{VDN significantly outperformed MAPPO} ($p = 0.0052$), highlighting the advantage of CTDE. Other comparisons, such as MAPPO vs.\ QMIX 
% or IPPO vs.\ IQL,
showed no significant difference.

\subsubsection{Impact of Team Composition on Performance.} 
Using the same test success rates metric, the results show that CTDE specifically, VDN consistently outperformed DTDE methods across team compositions, indicating strong generalizability to different team dynamics (see Table \ref{tab:combined_results}). 
VDN has the highest success rate in the uniform skilled teams in both tasks. 
However, the forced cooperation teams had the lowest success rates across all algorithms, reflecting the coordination burden imposed by asymmetric skill levels of agents. 
This performance drop highlights the challenge of learning robust policies when agents are unable to act interchangeably.

\subsubsection{Impact of Energy Function on Performance.} 
\begin{figure*}[t]
    \centering
    \includegraphics[width=0.7\textwidth]{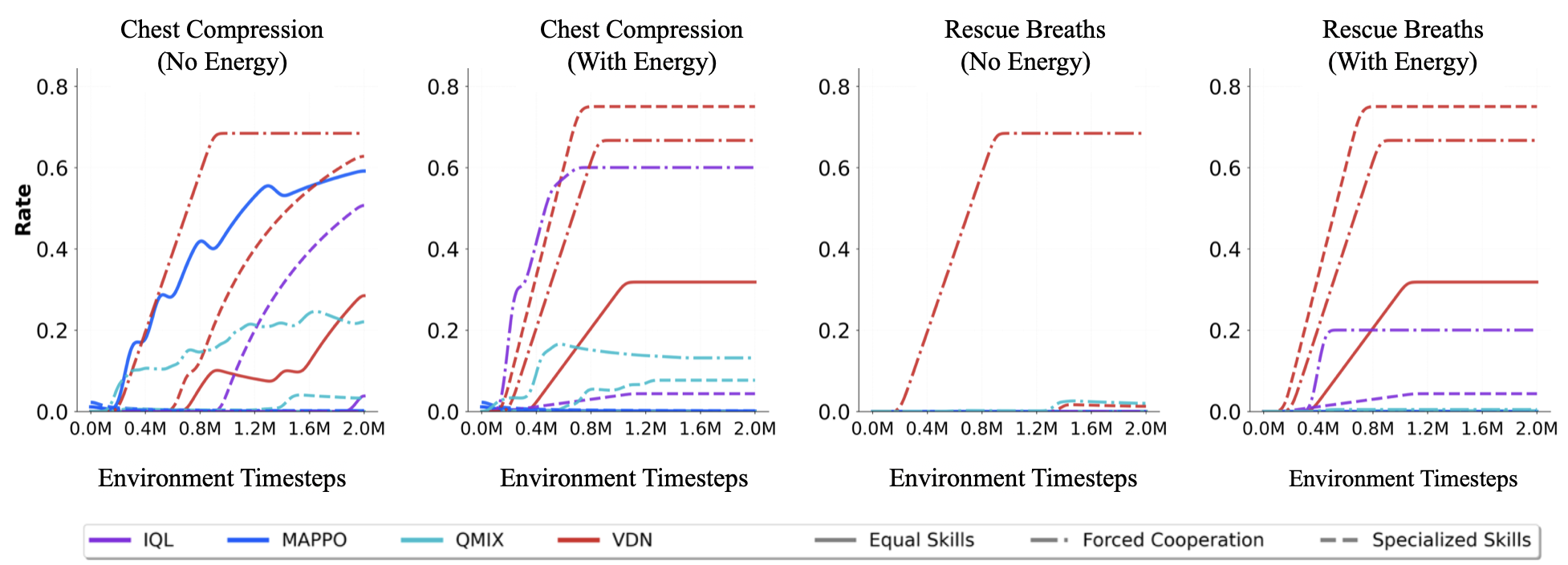} 
    \caption{Test cumulative rates for the ``chest compression" and ``rescue breaths" tasks, with and without energy.}
    \label{fig:all_scenarios_horizontal_comparison}
\end{figure*}

For the CPR goal, according to Table \ref{tab:combined_results}, the shared task mode (with an energy cost of  ``3” and recharge rate of  ``1” ) impacts the MARL algorithm’s performance compared to the baseline scenario (energy cost of  ``0”).
Under this setting, coordination becomes more challenging, as agents must alternate actions and manage limited energy budgets across a long-horizon task.
Despite this, the CTDE algorithm VDN shows the best performance in both the uniform (\textbf{PUE} = 0.84) and forced cooperation (\textbf{PFE} = 0.85) teams, outperforming all DTDE baselines. 
This suggests that centralized training may help agents better learn coordination strategies under resource constraints.
 In contrast, DTDE methods such as IQL %and IPPO
 struggle perhaps due to a lack of shared state.
MAPPO, despite struggling in the no energy team configurations achieves its best performance in the specialized setting with energy (\textbf{PSE} $=$ 0.97), indicating that role specialization might mitigate coordination difficulty  in this shared task setting.
Surprisingly across both tasks as seen in Figure \ref{fig:all_scenarios_horizontal_comparison}, agents tend to converge faster in the energy-constrained setting than in the no-energy baseline. This suggests that the structured turn-taking enforced by energy levels of agents in the shared task mode may simplify coordination rather than hinder performance, perhaps leading to more straightforward credit assignment and role specialization during learning.

% \usepackage{multirow}
\begin{table*}[t]
\centering
\caption{Success rates of MARL algorithms for CPR (P) at energy levels 0 and 3, and Rescue Breaths (C) at energy 0, with uniform (U), specialized (S), and forced cooperation (F) team compositions (higher is better).}
\begin{tabular}{l*{9}{c}}
\toprule
& \multicolumn{3}{c}{\textbf{Chest Compressions (Energy = 0)}} 
& \multicolumn{3}{c}{\textbf{Rescue Breaths  (Energy = 0)}} 
& \multicolumn{3}{c}{\textbf{Chest Compressions (Energy = 3)}} \\
\cmidrule(lr){2-4}\cmidrule(lr){5-7}\cmidrule(lr){8-10}
\textbf{Method} & P-U & P-S & P-F & C-U & C-S & C-F & P-U & P-S & P-F \\
\midrule
IQL \cite{tampuu_multiagent_2015}                                                      & 0.64           & {\color[HTML]{000000} 0.21}           & 0.51           & 0.46                  & 0.00                 & 0.32                 & 0.62 & 0.80                                            &   0.58                                            \\
% IPPO \cite{witt_is_2020}                                                     & 0.34           & {\color[HTML]{000000} 0.33}           & 0.64           & 0.11                  & 0.01                 & 0.05                 & 0.11 & 0.29                                            & 0.76                                            \\
MAPPO \cite{yu_surprising_2022}                                                    & 0.15           & {\color[HTML]{000000} 0.02}           & 0.02           & 0.01                  & 0.01                 & 0.00                 & 0.70 & \textbf{0.97 }                                           &   0.57                                            \\
VDN \cite{sunehag_value-decomposition_2017}                                                      & \textbf{0.85}  & {\color[HTML]{000000} \textbf{0.79}}  & \textbf{0.70}  & \textbf{0.80}                  & \textbf{0.74}                 & \textbf{0.61}                 & \textbf{0.84} & 0.78                                            &   \textbf{0.85}                                            \\
QMIX \cite{rashid_qmix_2018}                                                     & 0.84           & {\color[HTML]{000000} 0.68}           & 0.54           & 0.78                  & 0.60                 & 0.40                 & 0.79 & 0.60                                            &   0.63                                            \\ \hline
\end{tabular}
\label{tab:combined_results}
\end{table*}

\section{Experiment 2: FairSkillMARL}

\subsection{Baseline Methods}
\label{sec:fair_baseline}

We compare FairSkillMARL to two standard fairness methods across 500K timesteps and 4 seeds to investigate how well our framework performs in terms of skill-task alignment, workload, and both. \textbf{Gini Index} is a derivative of FairSkillMARL with $\alpha=1$, thus, it measures agents' contribution to sub-tasks in the reward function \cite{busa-fekete_multi-objective_2017, malfa_fairness_2025}. \textbf{Fair Efficient Network (FEN)} measures agent resource utilization and penalizes agents when they deviate from the average utilization of all agents \cite{jiang_learning_2019-1}.

\subsection{Training and Testing Procedure}
\label{sec:fair_train_test}

We conduct two controlled experiments across 500K timesteps and 4 seeds to isolate the effects of different fairness components. 
First, to isolate the impact of $L_2$,  we compared skill-task alignment ($\alpha = 0.7$) against workload only fairness ($\alpha = 0$) across four fairness weights ($\lambda \in \{5, 50, 100, 300\}$) using specialized teams due to their varying skill levels. 
% Second, to verify that the improved fairness is a result of skill-task alignment, not the total team skill, we compare equal and specialized teams at fixed $\lambda = 1.0$ with $\alpha \in \{0.0, 0.7\}$. 
Second, we compare all on-policy and off-policy MARL algorithms from Experiment 1 to evaluate fair task distributions in terms of workload and skill-alignment using the efficiency-based reward function, R($s_t, \mathbf{a}_t$).We used the specialised team setting with 3 agents for the rescue breaths task. All agents’ energy levels were set to 0; hence, the shared task mode was not explored.

\subsection{Measures}
\label{sec:fair_measures}

In addition to \textbf{success rate}, we use Workload balance and Alignment score as performance metrics to assess agents ability to balance workload and skill-task alignment as they aim to reach the goal. 

\textbf{Alignment score} measures the degree to which agents perform tasks aligned with their skill level, with 1.0 that indicates perfect alignment and 0 indicating random assignment. 
Specifically, it is the mean, over tasks executed in an episode, of the fraction performed by the task’s most-skilled agent (1 = perfect specialist use versus 0 = random). 
To the best of our knowledge, no prior MARL work quantifies fairness as the average fraction of tasks executed by the most‑skilled agent; the alignment score is therefore a novel contribution of this paper. See supplementary for mathematical formulation.

\textbf{Range} Workload balance is evaluated via the \emph{range} of normalized contributions, defined as the difference between the maximum and minimum agent contributions.  The range is the extent of workload distribution across agents, with 0 corresponding to perfectly balanced task sharing, while 1 reveals disproportionate burden. Range is a standard measure of workload equity in scheduling and routing research \cite{nielsen2024tactical, wolbeck2019fairness}. Meanwhile, the normalised contributions are calculated via the Gini index, which measures average pairwise disparity in contributions normalized by the mean. Gini‑based fairness metrics have been applied in MARL settings such as stock‑trading \cite{noauthor_fairness_nodate} and traffic‑signal control\cite{siddique_fairness_nodate}.  Combining these workload metrics with the alignment score provides a comprehensive assessment of fairness.

\subsection{FairSkillMARL Results}

\begin{figure}[t]
    \centering
    \includegraphics[width=0.5\textwidth]{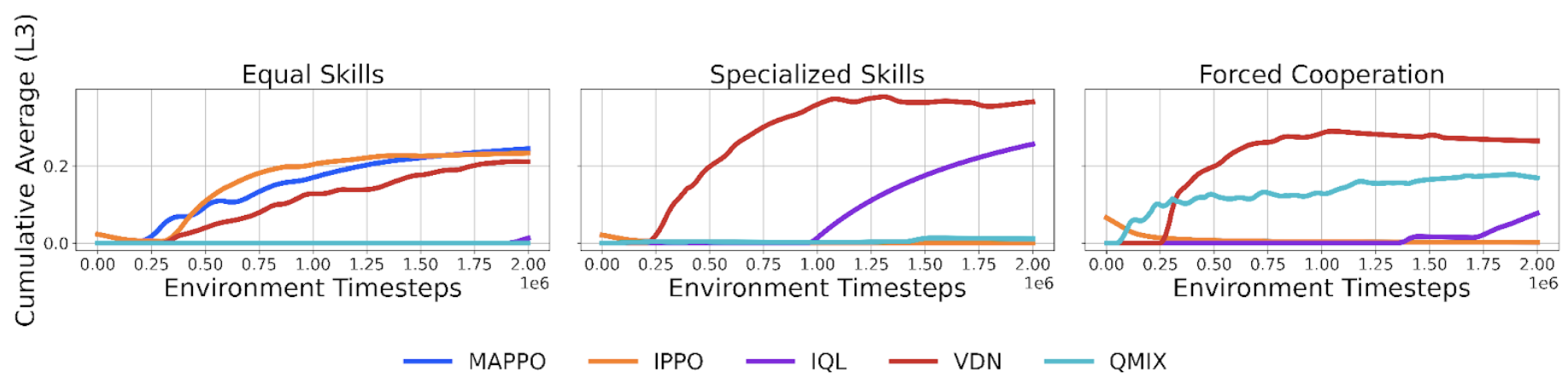} 
    \caption{$L_3$ composite disparity over training timesteps for different algorithms and skill configurations with no energy. Lower values indicate fairer agent behavior.}
    \label{fig:WO_cumulative_fairness}
\end{figure}

% \usepackage{multirow}
\begin{table*}[t]
\caption{Ablation study results to show the impact of skill-task alignment on specialized teams. Higher alignment and lower range indicate better coordination and fairness.}
\label{tab:fairness_results}
\begin{tabular}{llllll}
\hline
\textbf{Method}           & \textbf{$\lambda$}      & \textbf{$\alpha$} & \textbf{Success Rate $\uparrow$} & \textbf{Alignment Score $\uparrow$} & \textbf{Range $\downarrow$}       \\ \hline
FEN\cite{jiang_learning_2019}                       & \multirow{4}{*}{5}   & -              & 0.80 ± 0.02           & 0.32 ± 0.07              & 0.18 ± 0.09          \\
QMIX Gini \cite{malfa_fairness_2025}                 &                      & 1.0            & 0.82 ± 0.06           & 0.33 ± 0.08              & 0.06 ± 0.01          \\
QMIX FairSkillMARL (Ours) &                      & 0.0            & 0.80 ± 0.05           & \textbf{0.46 ± 0.19}     & \textbf{0.07 ± 0.03} \\
QMIX FairSkillMARL (Ours) &                      & 0.7            & \textbf{0.83 ± 0.04}  & 0.32 ± 0.16              & 0.06 ± 0.01          \\ \hline
FEN \cite{jiang_learning_2019}                       & \multirow{4}{*}{50}  & -              & \textbf{0.86 ± 0.10}  & 0.29 ± 0.03              & \textbf{0.06 ± 0.00} \\
QMIX Gini \cite{malfa_fairness_2025}                  &                      & 1.0            & 0.80 ± 0.03           & 0.42 ± 0.12              & 0.08 ± 0.05          \\
QMIX FairSkillMARL (Ours) &                      & 0.0            & 0.82 ± 0.04           & \textbf{0.48 ± 0.13}     & 0.10 ± 0.06          \\
QMIX FairSkillMARL (Ours) &                      & 0.7            & 0.83 ± 0.05           & 0.46 ± 0.18              & \textbf{0.06 ± 0.01} \\ \hline
FEN \cite{jiang_learning_2019}                       & \multirow{4}{*}{100} & -              & 0.40 ± 0.06           & 0.08 ± 0.03              & 0.36 ± 0.02          \\
QMIX Gini \cite{malfa_fairness_2025}                   &                      & 1.0            & \textbf{0.83±0.04}                     & 0.40±0.10                        & 0.17±0.07                    \\
QMIX FairSkillMARL (Ours) &                      & 0.0            & 0.80 ± 0.00  & \textbf{0.56 ± 0.00}     & \textbf{0.05 ± 0.00} \\
QMIX FairSkillMARL (Ours) &                      & 0.7            & 0.76 ± 0.00           & 0.22 ± 0.00              & 0.10 ± 0.00          \\ \hline
FEN \cite{jiang_learning_2019}                      & \multirow{4}{*}{300} & -              & 0.40 ± 0.06           & 0.10 ± 0.02              & 0.37 ± 0.01          \\
QMIX Gini \cite{jiang_learning_2019}                &                      & 1.0            & \textbf{0.83 ± 0.00}  & 0.23 ± 0.00              & 0.10 ± 0.00          \\
QMIX FairSkillMARL (Ours) &                      & 0.0            & 0.82 ± 0.06           & 0.41 ± 0.11              & \textbf{0.05 ± 0.01} \\
QMIX FairSkillMARL (Ours) &                      & 0.7            & 0.76 ± 0.00           & \textbf{0.55 ± 0.00}     & \textbf{0.05 ± 0.00} \\ \hline
\end{tabular}
\end{table*}

\subsubsection{\textbf{Comparison to Workload only baselines (GINI and FEN)}} We evaluate the effect of skill-task alignment ($L_2$) by comparing FairSkillMARL ($\alpha$ = 0.7) performance with workload-only baselines (QMIX Gini, $\alpha$ = 1.0 and FEN) across fairness weights $\lambda \in \{5, 50, 100, 300\}$. 
Our results from Table \ref{tab:fairness_results} show that across all conditions, FairSkillMARL achieves higher alignment scores while maintaining comparable or superior success rates.
A higher alignment score indicates that agents more frequently perform tasks matching their specialized skills (better coordination), whereas a lower workload range reflects more balanced task sharing across the team. \newline
Across fairness coefficients (\(\lambda = 5\)–\(300\)), \textit{FairSkillMARL} (\(\alpha = 0.7\)) consistently outperforms workload-only baselines (\textit{FEN} and \textit{QMIX Gini}) in both coordination quality and fairness.
At moderate fairness weights (\(\lambda = 5, 50\)), \textit{FairSkillMARL} improves alignment by \textbf{40--60\%} relative to \textit{FEN} and by \textbf{15--25\%} over \textit{QMIX Gini}, while reducing workload range by roughly \textbf{50\%} (0.06 vs.\ 0.10--0.18). 
At higher fairness levels (\(\lambda = 100, 300\)), \textit{FairSkillMARL} achieves stronger specialization. Alignment scores are \textbf{40--600\% higher} than \textit{FEN} and \textbf{35--40\% higher} than \textit{QMIX Gini}, while workload range is reduced by \textbf{70--85\%}. 
Despite the stronger regularization, success rates remain comparable or higher (0.80--0.83 vs.\ 0.40--0.86), indicating that incorporating explicit skill--task alignment (\(L_2\)) enables teams to remain both fair and efficient even under strong fairness constraints. %As a sanity check, 
We expect (\(\alpha = 0.0\)) to have the highest alignment score, as the fairness reward is collapsed to only the $L_2$ skill-task alignment metric after that. For (\(\lambda = 5,50, 100\)), we observe that trend further confirming the impact of $L_2$.

These results demonstrate that including skill–task alignment (\(L_2\)) in the fairness definition produces fair teams with \textbf{better coordination}, unlike workload-only approaches, which focus on equalised efforts without considering agent specialization. \textbf{This robustness is most evident at extreme fairness ($\lambda=300$), where FEN degrades to 40\% success while FairSkillMARL maintains 76\% success, showing that skill-aware coordination remains effective where pure workload balancing may fail.}
Interestingly, in some settings (\(\lambda = 5, 50, 100\)), \(\alpha = 0\) yields higher alignment and lower range because the fairness reward (\(L_3 = \alpha L_1 + (1 - \alpha)L_2\)) focuses solely on skill--task alignment, leading to more coordinated behavior and efficient completion of tasks. Meanwhile, higher \(\alpha\) (e.g., 0.7) prioritizes workload balance at the cost of specialization.

\subsubsection{\textbf{Comparison Across Algorithms}} We investigated the emergent fairness of standard MARL algorithms without 
fairness reward shaping. IQL and VDN exhibit severe agent imbalance, while MAPPO achieves balanced coordination. Under energy constraints, MAPPO remains robust, 
%IPPO underutilizes agents,
and value-based methods (IQL, VDN, QMIX) continue to show imbalance as seen in Figure \ref{fig:WO_cumulative_fairness}.

\section{Experiment 3: FairSkillMARL Ablation}
\label{sec:exp-rff2}

\subsection{Baseline Methods, Training, and Testing Procedure}
\label{sec:fair_train_teststatsig}

We compared FairskillMARL variants from Experiment \ref{sec:fair_baseline} against workload baselines (FEN and QMIX Gini) at moderate fairness parameter $\lambda = 50$ across 500k timesteps with 6 seeds.
% , using Mann-Whitney U tests, a non-parametric method robust to non-normal distributions common in RL experiments. 
Our specialised team setting here has all agents' energy levels set to 0, hence the shared task mode was not explored.

\subsection{Measures}
\label{sec:fair_measuresstatsign}
Our experimental design involves three independent variables (Algorithm type, fairness penalty magnitude ($\lambda$), and random seeds. Primary dependent variables were success rate and skill-task alignment score. We use Mann-Whitney U tests, a non-parametric method robust to non-normal distributions common in RL experiments. 
\subsection{Statistical Significance Results}
When configured to prioritize skill-task alignment ($\alpha$=0.0), FairSkillMARL achieves marginally significantly higher alignment scores than FEN (0.48 vs 0.29, p=0.095, effect size r=0.47), implying that imposing the skill-task fairness penalty can improve agent's skill alignment to their tasks.
At $\alpha=1.0$, FairSkillMARL behaves similarly to QMIX Gini,showing that our framework includes existing fairness methods as a special case. This demonstrates that our $\alpha$ parameter correctly controls the fairness objective.
% FairSkillMARL ($\alpha$=0.7) demonstrated statistically significantly higher success rates compared to the FEN baseline (Mann-Whitney U = 103, p = 0.010, r = 0.53). 

% FairSkillMARL achieved a mean success rate of 77.1\% (SD = 7.0\%) compared to 70.6\% achieved by FEN (SD = 7.4\%), representing a 6.5\%  improvement with a large effect size. 

\section{Conclusion}

In this work, we introduced skill-task alignment as a fairness criterion for heterogeneous agents in MARL. We also introduced MARLHospital, a customizable simulation environment for benchmarking cooperative MARL algorithms on varying skill levels and energy levels across task difficulty levels. We evaluated four standard MARL algorithms across two tasks with three agents with different skill levels and energy cost constraints. %We evaluated the effectiveness of IL and CTDE algorithms across different skill levels and energy cost constraints, finding 
We found that CTDE algorithms achieved the best performance in both the CPR task and the rescue breaths tasks. 
Also, our experiments show that FairSkillMARL achieves a statistically significant improvement at $\alpha$=0.7 compared to the FEN algorithm, and that our approach captures skill misalignment in complex coordination tasks.% might require more robust metrics. %Our experiments with FairSkillMARL in MARLHospital are informative, achieving statistical improvements at $\alpha$=0.7 compared to pure workload in the FEN algorithm, capturing skill misalignment in complex coordination tasks might require more robust metrics. Our findings suggest that FairSkillMARL does not incur a performance drop while optimizing for skill-task alignment.
 By modeling agent capabilities explicitly, our method provides a foundation for fairness definitions that consider both workload and agent skills in heterogeneous teams. Future research can build upon this foundation to explore multi-objective optimization techniques, and applications to larger-scale heterogeneous multi-agent systems.

\bibliographystyle{unsrt}  
\bibliography{templateArxiv2_clean}  

\begin{thebibliography}{10}

\bibitem{taylor_rapidly_2025}
Angelique Taylor, Tauhid Tanjim, Michael~Joseph Sack, Maia Hirsch, Kexin Cheng, Kevin Ching, Jonathan~St George, Thijs Roumen, Malte~F. Jung, and Hee~Rin Lee.
\newblock Rapidly {Built} {Medical} {Crash} {Cart}! {Lessons} {Learned} and {Impacts} on {High}-{Stakes} {Team} {Collaboration} in the {Emergency} {Room}, February 2025.
\newblock arXiv:2502.18688 [cs] version: 1.

\bibitem{taylor_towards_2024}
Angelique Taylor, Tauhid Tanjim, Huajie Cao, and Hee~Rin Lee.
\newblock Towards {Collaborative} {Crash} {Cart} {Robots} that {Support} {Clinical} {Teamwork}.
\newblock In {\em Proceedings of the 2024 {ACM}/{IEEE} {International} {Conference} on {Human}-{Robot} {Interaction}}, {HRI} '24, pages 715--724, New York, NY, USA, March 2024. Association for Computing Machinery.

\bibitem{taylor_coordinating_2019}
Angelique Taylor, Hee~Rin Lee, Alyssa Kubota, and Laurel~D. Riek.
\newblock Coordinating {Clinical} {Teams}: {Using} {Robots} to {Empower} {Nurses} to {Stop} the {Line}.
\newblock {\em Proceedings of the ACM on Human-Computer Interaction}, 3(CSCW):1--30, November 2019.

\bibitem{tampuu_multiagent_2015}
Ardi Tampuu, Tambet Matiisen, Dorian Kodelja, Ilya Kuzovkin, Kristjan Korjus, Juhan Aru, Jaan Aru, and Raul Vicente.
\newblock Multiagent {Cooperation} and {Competition} with {Deep} {Reinforcement} {Learning}, November 2015.

\bibitem{witt_is_2020}
Christian Schroeder~de Witt, Tarun Gupta, Denys Makoviichuk, Viktor Makoviychuk, Philip H.~S. Torr, Mingfei Sun, and Shimon Whiteson.
\newblock Is {Independent} {Learning} {All} {You} {Need} in the {StarCraft} {Multi}-{Agent} {Challenge}?, November 2020.

\bibitem{sunehag_value-decomposition_2017}
Peter Sunehag, Guy Lever, Audrunas Gruslys, Wojciech~Marian Czarnecki, Vinicius Zambaldi, Max Jaderberg, Marc Lanctot, Nicolas Sonnerat, Joel~Z. Leibo, Karl Tuyls, and Thore Graepel.
\newblock Value-{Decomposition} {Networks} {For} {Cooperative} {Multi}-{Agent} {Learning}, June 2017.

\bibitem{rashid_qmix_2018}
Tabish Rashid, Mikayel Samvelyan, Christian~Schroeder de~Witt, Gregory Farquhar, Jakob Foerster, and Shimon Whiteson.
\newblock {QMIX}: {Monotonic} {Value} {Function} {Factorisation} for {Deep} {Multi}-{Agent} {Reinforcement} {Learning}, June 2018.

\bibitem{carroll_utility_2020}
Micah Carroll, Rohin Shah, Mark~K. Ho, Thomas~L. Griffiths, Sanjit~A. Seshia, Pieter Abbeel, and Anca Dragan.
\newblock On the {Utility} of {Learning} about {Humans} for {Human}-{AI} {Coordination}, January 2020.

\bibitem{wang_demo2code_2023}
Huaxiaoyue Wang, Gonzalo Gonzalez-Pumariega, Yash Sharma, and Sanjiban Choudhury.
\newblock {Demo2Code}: {From} {Summarizing} {Demonstrations} to {Synthesizing} {Code} via {Extended} {Chain}-of-{Thought}, November 2023.

\bibitem{samvelyan_starcraft_2019}
Mikayel Samvelyan, Tabish Rashid, Christian Schroeder~de Witt, Gregory Farquhar, Nantas Nardelli, Tim G.~J. Rudner, Chia-Man Hung, Philip H.~S. Torr, Jakob Foerster, and Shimon Whiteson.
\newblock The {StarCraft} {Multi}-{Agent} {Challenge}, December 2019.

\bibitem{ellis_smacv2_2023}
Benjamin Ellis, Jonathan Cook, Skander Moalla, Mikayel Samvelyan, Mingfei Sun, Anuj Mahajan, Jakob~N. Foerster, and Shimon Whiteson.
\newblock {SMACv2}: {An} {Improved} {Benchmark} for {Cooperative} {Multi}-{Agent} {Reinforcement} {Learning}, October 2023.

\bibitem{hughes_inequity_2018}
Edward Hughes, Joel~Z. Leibo, Matthew~G. Phillips, Karl Tuyls, Edgar~A. Duéñez-Guzmán, Antonio~García Castañeda, Iain Dunning, Tina Zhu, Kevin~R. McKee, Raphael Koster, Heather Roff, and Thore Graepel.
\newblock Inequity aversion improves cooperation in intertemporal social dilemmas, September 2018.
\newblock arXiv:1803.08884 [cs].

\bibitem{yuan_multi-agent_2021}
Mingqi Yuan, Qi~Cao, Man-On Pun, and Yi~Chen.
\newblock Multi-{Agent} {Reinforcement} {Learning}-{Based} {Fairness}-{Aware} {Scheduling} for {Bursty} {Traffic}.
\newblock In {\em 2021 {IEEE} {Global} {Communications} {Conference} ({GLOBECOM})}, pages 1--6, December 2021.

\bibitem{fang_fairness-aware_2024}
Wanqing Fang, Xintian Zhao, and Chengwei Zhang.
\newblock Fairness-aware multi-agent reinforcement learning and visual perception for adaptive traffic signal control.
\newblock {\em Optoelectronics Letters}, 20(12):764--768, December 2024.

\bibitem{huang_fairness-aware_2022}
Xingshuai Huang, Di~Wu, and Benoit Boulet.
\newblock Fairness-{Aware} {Model}-{Based} {Multi}-{Agent} {Reinforcement} {Learning} for {Traffic} {Signal} {Control}.
\newblock September 2022.

\bibitem{aloor_cooperation_2024}
Jasmine~Jerry Aloor, Siddharth Nayak, Sydney Dolan, and Hamsa Balakrishnan.
\newblock Cooperation and {Fairness} in {Multi}-{Agent} {Reinforcement} {Learning}, October 2024.

\bibitem{jiang_learning_2019}
Jiechuan Jiang and Zongqing Lu.
\newblock Learning {Fairness} in {Multi}-{Agent} {Systems}, October 2019.
\newblock arXiv:1910.14472 [cs].

\bibitem{malfa_fairness_2025}
Gabriele~La Malfa, Jie~M. Zhang, Michael Luck, and Elizabeth Black.
\newblock Fairness {Aware} {Reinforcement} {Learning} via {Proximal} {Policy} {Optimization}, September 2025.
\newblock arXiv:2502.03953 [cs].

\bibitem{li_agent_2024}
Junkai Li, Siyu Wang, Meng Zhang, Weitao Li, Yunghwei Lai, Xinhui Kang, Weizhi Ma, and Yang Liu.
\newblock Agent {Hospital}: {A} {Simulacrum} of {Hospital} with {Evolvable} {Medical} {Agents}, May 2024.

\bibitem{scheikl_cooperative_2021}
Paul~Maria Scheikl, Balázs Gyenes, Tornike Davitashvili, Rayan Younis, André Schulze, Beat~P. Müller-Stich, Gerhard Neumann, Martin Wagner, and Franziska Mathis-Ullrich.
\newblock Cooperative {Assistance} in {Robotic} {Surgery} through {Multi}-{Agent} {Reinforcement} {Learning}.
\newblock In {\em 2021 {IEEE}/{RSJ} {International} {Conference} on {Intelligent} {Robots} and {Systems} ({IROS})}, pages 1859--1864, September 2021.
\newblock arXiv:2110.04857 [cs].

\bibitem{zhang_multi-agent_2024}
Xinzhi Zhang, Yeming Yang, Qingling Zhu, Qiuzhen Lin, Weineng Chen, Jianqiang Li, and Carlos A.~Coello Coello.
\newblock Multi-agent deep {Q}-network-based metaheuristic algorithm for {Nurse} {Rostering} {Problem}.
\newblock {\em Swarm and Evolutionary Computation}, 87:101547, June 2024.

\bibitem{gong_mindagent_2023}
Ran Gong, Qiuyuan Huang, Xiaojian Ma, Hoi Vo, Zane Durante, Yusuke Noda, Zilong Zheng, Song-Chun Zhu, Demetri Terzopoulos, Li~Fei-Fei, and Jianfeng Gao.
\newblock {MindAgent}: {Emergent} {Gaming} {Interaction}, September 2023.

\bibitem{puig_virtualhome_2018}
Xavier Puig, Kevin Ra, Marko Boben, Jiaman Li, Tingwu Wang, Sanja Fidler, and Antonio Torralba.
\newblock {VirtualHome}: {Simulating} {Household} {Activities} via {Programs}, June 2018.

\bibitem{resnick_pommerman_2018}
Cinjon Resnick, Wes Eldridge, David Ha, Denny Britz, Jakob Foerster, Julian Togelius, Kyunghyun Cho, and Joan Bruna.
\newblock Pommerman: {A} {Multi}-{Agent} {Playground}, September 2018.
\newblock Publication Title: arXiv.org.

\bibitem{agapiou_melting_2023}
John~P. Agapiou, Alexander~Sasha Vezhnevets, Edgar~A. Duéñez-Guzmán, Jayd Matyas, Yiran Mao, Peter Sunehag, Raphael Köster, Udari Madhushani, Kavya Kopparapu, Ramona Comanescu, D.~J. Strouse, Michael~B. Johanson, Sukhdeep Singh, Julia Haas, Igor Mordatch, Dean Mobbs, and Joel~Z. Leibo.
\newblock Melting {Pot} 2.0, October 2023.

\bibitem{hao_hierarchical_2023}
Qianyue Hao, Fengli Xu, Lin Chen, Pan Hui, and Yong Li.
\newblock Hierarchical {Multi}-agent {Model} for {Reinforced} {Medical} {Resource} {Allocation} with {Imperfect} {Information}.
\newblock {\em ACM Transactions on Intelligent Systems and Technology}, 14(1):1--27, February 2023.

\bibitem{esposito_multi-agent_2020}
Dario Esposito, Davide Schaumann, Domenico Camarda, and Yehuda~E. Kalay.
\newblock Multi-{Agent} {Modelling} and {Simulation} of {Hospital} {Acquired} {Infection} {Propagation} {Dynamics} by {Contact} {Transmission} in {Hospital} {Wards}.
\newblock In Yves Demazeau, Tom Holvoet, Juan~M. Corchado, and Stefania Costantini, editors, {\em Advances in {Practical} {Applications} of {Agents}, {Multi}-{Agent} {Systems}, and {Trustworthiness}. {The} {PAAMS} {Collection}}, pages 118--133, Cham, 2020. Springer International Publishing.

\bibitem{luo_h2-marl_2025}
Xueting Luo, Hao Deng, Jihong Yang, Yao Shen, Huanhuan Guo, Zhiyuan Sun, Mingqing Liu, Jiming Wei, and Shengjie Zhao.
\newblock H2-{MARL}: {Multi}-{Agent} {Reinforcement} {Learning} for {Pareto} {Optimality} in {Hospital} {Capacity} {Strain} and {Human} {Mobility} during {Epidemic}, March 2025.

\bibitem{koduri_nurseschedrl_2025}
Harsha Koduri.
\newblock {NurseSchedRL}: {Attention}-{Guided} {Reinforcement} {Learning} for {Nurse}-{Patient} {Assignment}, September 2025.
\newblock arXiv:2509.18125 [cs].

\bibitem{ruiz_mejia_medscrubcrew_2025}
Jose~M. Ruiz~Mejia and Danda~B. Rawat.
\newblock {MedScrubCrew}: {A} {Medical} {Multi}-{Agent} {Framework} for {Automating} {Appointment} {Scheduling} {Based} on {Patient}-{Provider} {Profile} {Resource} {Matching}.
\newblock {\em Healthcare}, 13(14):1649, July 2025.
\newblock Publisher: Multidisciplinary Digital Publishing Institute.

\bibitem{torres-huesca_multi-agent_2021}
Angélica Torres-Huesca, Luis-Gerardo Montané-Jiménez, and María-Teresa Cepero-García.
\newblock {MULTI}-{AGENT} {COLLABORATIVE} {WORK} {SIMULATION} {FOR} {PATIENT} {CARE} {IN} {HOSPITALS}.
\newblock {\em DYNA New Technologies Journal}, 8(1), 2021.

\bibitem{amato_first_2024}
Christopher Amato.
\newblock A {First} {Introduction} to {Cooperative} {Multi}-{Agent} {Reinforcement} {Learning}, December 2024.

\bibitem{schulman_proximal_2017}
John Schulman, Filip Wolski, Prafulla Dhariwal, Alec Radford, and Oleg Klimov.
\newblock Proximal {Policy} {Optimization} {Algorithms}, August 2017.

\bibitem{yu_surprising_2022}
Chao Yu, Akash Velu, Eugene Vinitsky, Jiaxuan Gao, Yu~Wang, Alexandre Bayen, and Yi~Wu.
\newblock The {Surprising} {Effectiveness} of {PPO} in {Cooperative}, {Multi}-{Agent} {Games}, November 2022.

\bibitem{foerster_counterfactual_2018}
Jakob Foerster, Gregory Farquhar, Triantafyllos Afouras, Nantas Nardelli, and Shimon Whiteson.
\newblock Counterfactual {Multi}-{Agent} {Policy} {Gradients}.
\newblock {\em Proceedings of the AAAI Conference on Artificial Intelligence}, 32(1), April 2018.

\bibitem{perera_demystifying_2023}
Dilruk Perera, Siqi Liu, and Mengling Feng.
\newblock Demystifying {Complex} {Treatment} {Recommendations}: {A} {Hierarchical} {Cooperative} {Multi}-{Agent} {RL} {Approach}.
\newblock In {\em 2023 {International} {Joint} {Conference} on {Neural} {Networks} ({IJCNN})}, pages 1--10, June 2023.

\bibitem{tan_advancing_2024}
Daniel~J. Tan, Qianyi Xu, Kay~Choong See, Dilruk Perera, and Mengling Feng.
\newblock Advancing {Multi}-{Organ} {Disease} {Care}: {A} {Hierarchical} {Multi}-{Agent} {Reinforcement} {Learning} {Framework}, September 2024.

\bibitem{guss_minerl_2019}
William~H. Guss, Brandon Houghton, Nicholay Topin, Phillip Wang, Cayden Codel, Manuela Veloso, and Ruslan Salakhutdinov.
\newblock {MineRL}: {A} {Large}-{Scale} {Dataset} of {Minecraft} {Demonstrations}, July 2019.

\bibitem{gonzalez-pumariega_robotouille_2025}
Gonzalo Gonzalez-Pumariega, Leong~Su Yean, Neha Sunkara, and Sanjiban Choudhury.
\newblock Robotouille: {An} {Asynchronous} {Planning} {Benchmark} for {LLM} {Agents}, February 2025.
\newblock arXiv:2502.05227 [cs].

\bibitem{bredereck_computing_2025}
Robert Bredereck.
\newblock Computing {Efficient} {Envy}-{Free} {Partial} {Allocations} of {Indivisible} {Goods}.
\newblock 2025.

\bibitem{barman_parameterized_2024}
Siddharth Barman, Debajyoti Kar, and Shraddha Pathak.
\newblock Parameterized {Guarantees} for {Almost} {Envy}-{Free} {Allocations}.
\newblock {\em New Zealand}, 2024.

\bibitem{mutzari_resilient_2023}
Dolev Mutzari.
\newblock Resilient {Fair} {Allocation} of {Indivisible} {Goods}.
\newblock 2023.

\bibitem{scarlett_for_2023}
Jonathan Scarlett.
\newblock For {One} and {All}: {Individual} and {Group} {Fairness} in the {Allocation} of {Indivisible} {Goods}.
\newblock 2023.

\bibitem{smit_fairness_2024}
Jacobus Smit.
\newblock Fairness and {Cooperation} between {Independent} {Reinforcement} {Learners} through {Indirect} {Reciprocity}.
\newblock {\em New Zealand}, 2024.

\bibitem{sood_fairness_2024}
Archit Sood.
\newblock Fairness of {Exposure} in {Online} {Restless} {Multi}-armed {Bandits}.
\newblock {\em New Zealand}, 2024.

\bibitem{siddique_towards_2025}
Umer Siddique.
\newblock Towards {Fair} and {Efficient} {Policy} {Learning} in {Cooperative} {Multi}-{Agent} {Reinforcement} {Learning}.
\newblock 2025.

\bibitem{papoudakis_benchmarking_2021}
Georgios Papoudakis, Filippos Christianos, Lukas Schäfer, and Stefano~V. Albrecht.
\newblock Benchmarking {Multi}-{Agent} {Deep} {Reinforcement} {Learning} {Algorithms} in {Cooperative} {Tasks}, November 2021.

\bibitem{oliehoek_concise_2016}
Frans Oliehoek and Christopher Amato.
\newblock A {Concise} {Introduction} to {Decentralized} {POMDPs}.
\newblock January 2016.

\bibitem{savatmongkorngul_comparison_nodate}
Sorravit Savatmongkorngul, Chaiyaporn Yuksen, Sumalin Chumkot, Pongsakorn Atiksawedparit, Chetsadakon Jenpanitpong, Sorawich Watcharakitpaisan, Parama Kaninworapan, and Konwachira Maijan.
\newblock Comparison of chest compression quality between 2-minute... : {International} {Journal} of {Critical} {Illness} and {Injury} {Science}.

\bibitem{foerster_learning_2016}
Jakob Foerster, Ioannis~Alexandros Assael, Nando de~Freitas, and Shimon Whiteson.
\newblock Learning to {Communicate} with {Deep} {Multi}-{Agent} {Reinforcement} {Learning}.
\newblock In {\em Advances in {Neural} {Information} {Processing} {Systems}}, volume~29. Curran Associates, Inc., 2016.

\bibitem{he_opponent_2016}
He~He, Jordan Boyd-Graber, Kevin Kwok, and Hal~Daumé III.
\newblock Opponent {Modeling} in {Deep} {Reinforcement} {Learning}, September 2016.

\bibitem{du_liir_2019}
Yali Du, Lei Han, Meng Fang, Ji~Liu, Tianhong Dai, and Dacheng Tao.
\newblock {LIIR}: {Learning} {Individual} {Intrinsic} {Reward} in {Multi}-{Agent} {Reinforcement} {Learning}.
\newblock In {\em Advances in {Neural} {Information} {Processing} {Systems}}, volume~32. Curran Associates, Inc., 2019.

\bibitem{mnih_playing_2013}
Volodymyr Mnih, Koray Kavukcuoglu, David Silver, Alex Graves, Ioannis Antonoglou, Daan Wierstra, and Martin Riedmiller.
\newblock Playing {Atari} with {Deep} {Reinforcement} {Learning}, December 2013.

\bibitem{busa-fekete_multi-objective_2017}
Robert Busa-Fekete, Balazs Szorenyi, Paul Weng, and Shie Mannor.
\newblock Multi-objective {Bandits}: {Optimizing} the {Generalized} {Gini} {Index}, June 2017.
\newblock arXiv:1706.04933 [cs].

\bibitem{jiang_learning_2019-1}
Jiechuan Jiang and Zongqing Lu.
\newblock Learning {Fairness} in {Multi}-{Agent} {Systems}, October 2019.
\newblock arXiv:1910.14472 [cs].

\bibitem{noauthor_fairness_nodate}
Fairness in {Multi}-agent {Reinforcement} {Learning} for {Stock} {Trading}.

\bibitem{siddique_fairness_nodate}
Umer Siddique, Peilang Li, and Yongcan Cao.
\newblock Fairness in {Traffic} {Control}: {Decentralized} {Multi}-agent {Reinforcement} {Learning} with {Generalized} {Gini} {Welfare} {Functions}.

\bibitem{mnih_human-level_2015}
Volodymyr Mnih, Koray Kavukcuoglu, David Silver, Andrei~A. Rusu, Joel Veness, Marc~G. Bellemare, Alex Graves, Martin Riedmiller, Andreas~K. Fidjeland, Georg Ostrovski, Stig Petersen, Charles Beattie, Amir Sadik, Ioannis Antonoglou, Helen King, Dharshan Kumaran, Daan Wierstra, Shane Legg, and Demis Hassabis.
\newblock Human-level control through deep reinforcement learning.
\newblock {\em Nature}, 518(7540):529--533, February 2015.

\end{thebibliography}

\clearpage   
\newpage
\appendix

\section{MARLHospital Environment}
\label{app: MARLHospital}

\FloatBarrier              % flush out earlier floats
\begin{figure}[!htb]
  \centering
  % in two-column mode, \columnwidth == one column’s width
  \includegraphics[width=0.9\columnwidth]{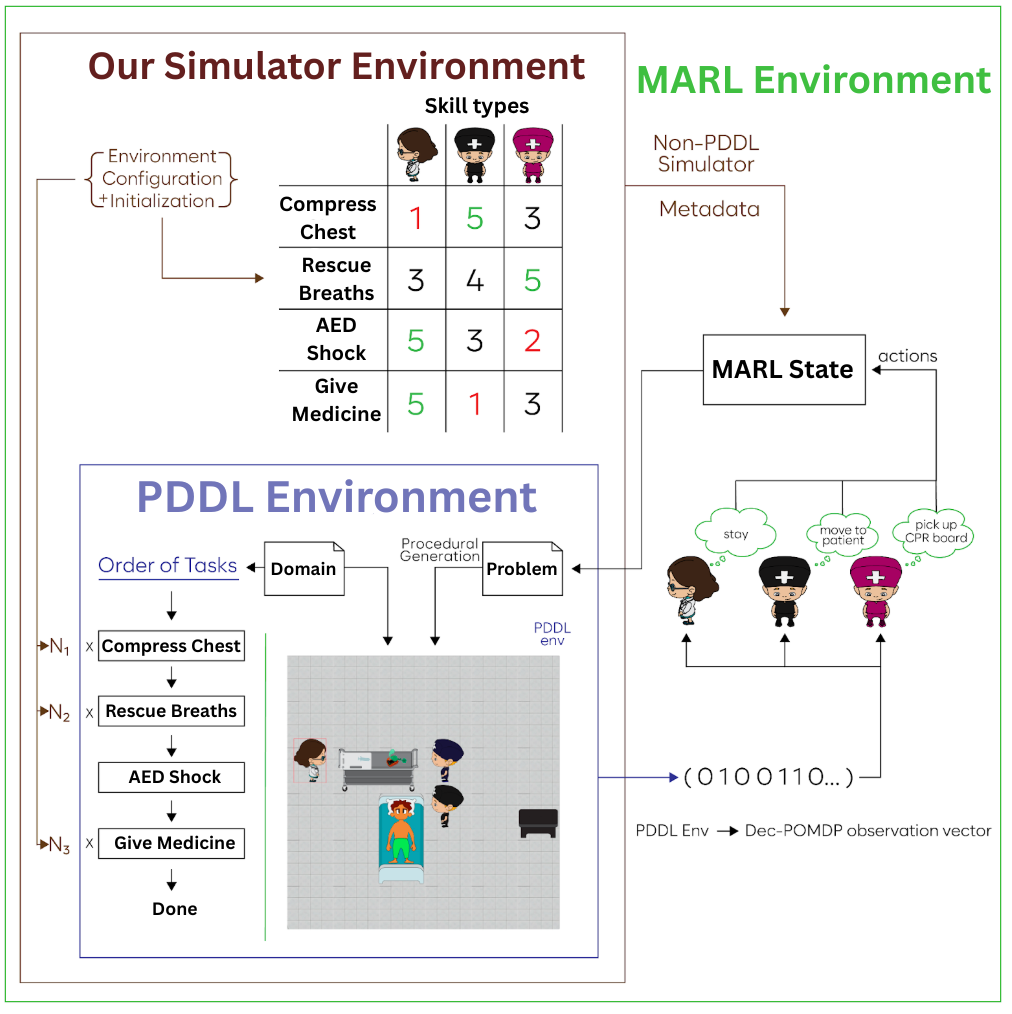} 
  \caption{MARLHospital environment: Breakdown of the PDDL environment wrapped by the MARL environment with 3 HCW agents and a patient.}
  \label{fig:main-figure-draft}
\end{figure}
\FloatBarrier              % prevent later text from floating above 

% \begin{figure}[h]
%     \includegraphics[width=0.7\textwidth]{images/Graphic illustration 3.png} 
%     \caption{MARLHospital environment: Breakdown of the PDDL environment wrapped by the MARL environment with 3 HCW agents and a patient.}
%     \label{fig: main figure draft}
% \end{figure}

Figure \ref{fig:main-figure-draft} shows how the Planning Domain Definition Language (PDDL) and MARL environment are integrated within MARLHospital in the baseline configuration. A configuration file initializes 1) the agent, station, patient, and object positions in the PDDL environment, 2) the skill (and possibly initial energy) levels of the agents, and 3) the subtask requirements to reach the goal. In the default setting, agents receive a binary encoding of the PDDL environment as an observation; when skill information is included in the observation, that data is passed from the configuration as well. The MARL state converts agents' actions into pddl actions based on state factors such as skill level, energy, and subtask requirements marked by $N_1$, $N_2$, and $N_3$ in the diagram.

The remainder of this section provides an overview of the different components of the PDDL environment.

\begin{figure}[h]
  \centering
  \includegraphics[width=0.1\textwidth]{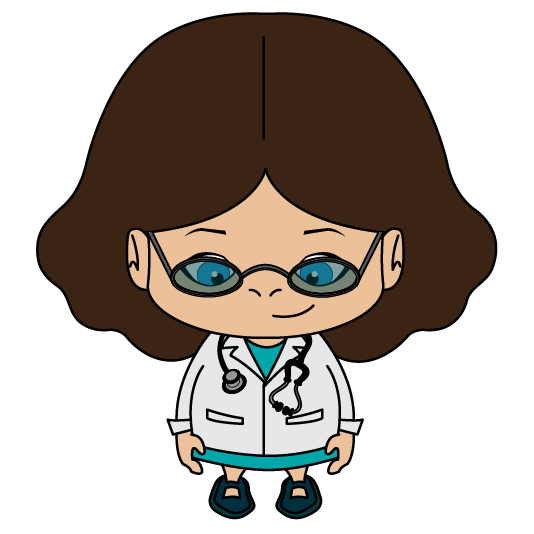}
  \includegraphics[width=0.1\textwidth]{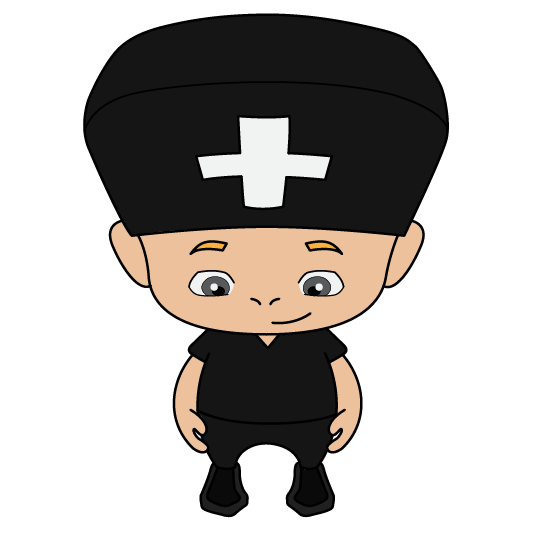}
  \includegraphics[width=0.1\textwidth]{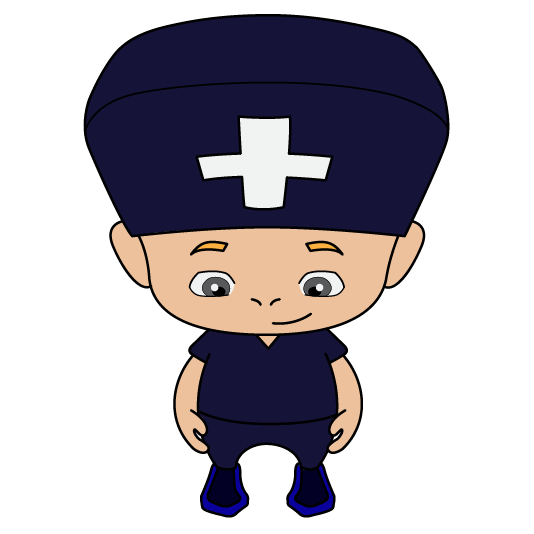}
  \includegraphics[width=0.1\textwidth]{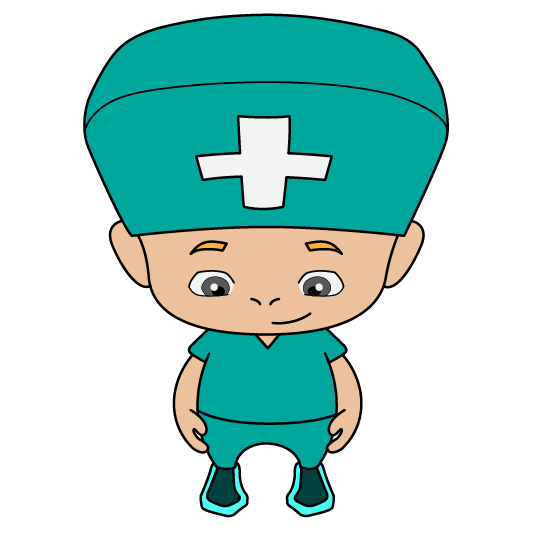}
  \includegraphics[width=0.1\textwidth]{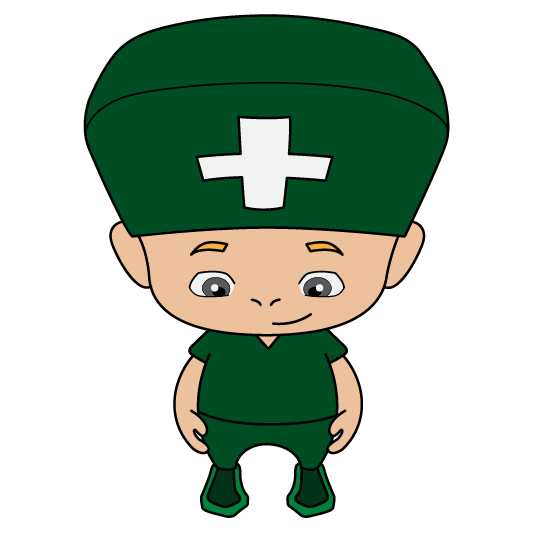}
  \includegraphics[width=0.1\textwidth]{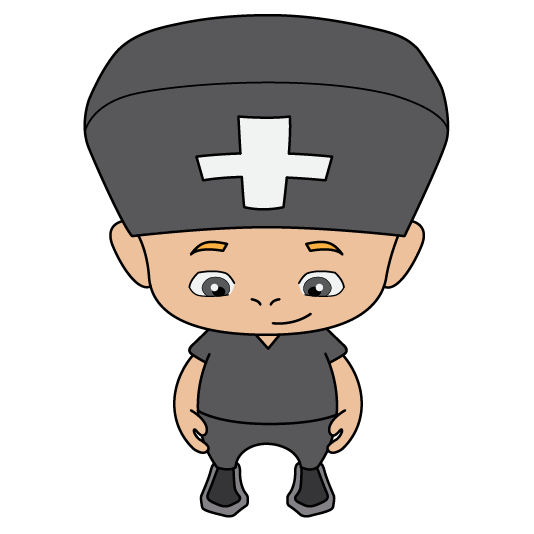}
  \includegraphics[width=0.1\textwidth]{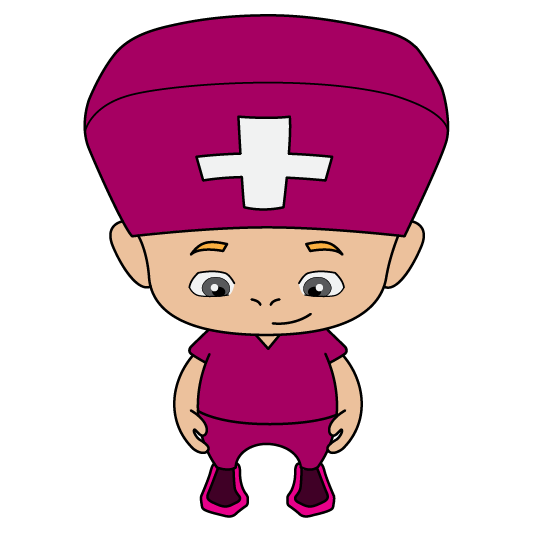}
  \caption{The PDDL players, controlled by RL agents}
  \label{fig:player_imgs}
\end{figure}

\textbf{Players} consist of the healthcare workers in the environment. Controlled by the RL agents, they take actions in the PDDL environment, and are allowed to move to locations on the grid adjacent to stations. See Figure \ref{fig:player_imgs} for player options.

\textbf{Actions} consist of the actions healthcare workers can perform within the environment. These include the following:

\begin{itemize}
    \item Move: move the worker from station A to station B
    \item Pick-up: pick up an item at the worker's current station. Afterwards, the station is empty and the worker now holds the item.
    \item Place: place the item the worker is holding on the worker's current station. Before placing an item, the station must be empty. Afterwards player no longer holds an item.
    \item Unstack: unstack the item that is on top of another item at the worker's current station. The unstacked item is now held by the worker.
    \item Stack: stack an item on top of another item at the worker's current station. The worker no longer holds an item.
    \item Stack under: stack an item underneath another item at the worker's current station.
    \item Chest compression: perform a medical procedure in which the worker manually circulates the blood of the patient. Before compressing chest, the patient must have a CPR board underneath them.
    \item Rescue breaths: perform a medical procedure in which the worker directly provides oxygen to the patient. Before giving rescue breaths, the patient must have a breathing pump attached.
    \item Give shock: perform a medical a medical procedure in which the worker uses a defibrillator to restart the patient's heart. The patient must have an AED device connected to them.
    \item Give medicine: perform a medical procedure in which the worker administers medicine to the patient through a syringe. The syringe must be on the patient.
    
\end{itemize}moving between stations, picking up or placing items, stacking or unstacking objects (including stacking under), and performing medical procedures such as chest compressions and rescue breaths.

% , defibrillation, and medication administration.

\begin{figure}[h]
  \centering
  \includegraphics[width=0.1\textwidth]{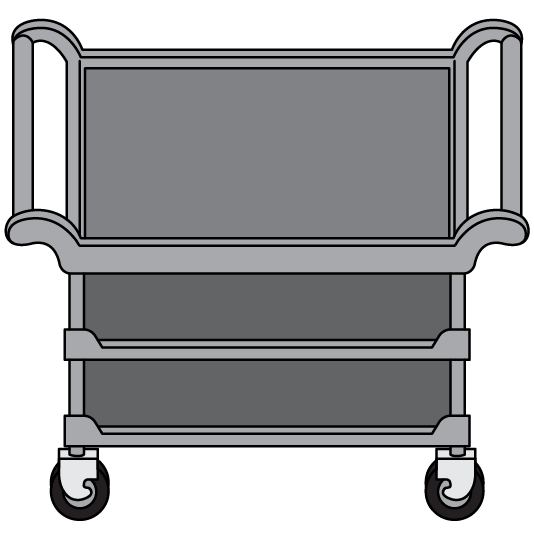}
  \includegraphics[width=0.1\textwidth]{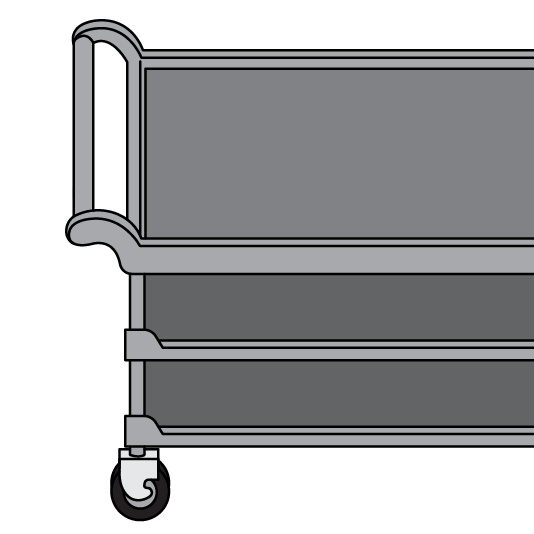}
  \includegraphics[width=0.1\textwidth]{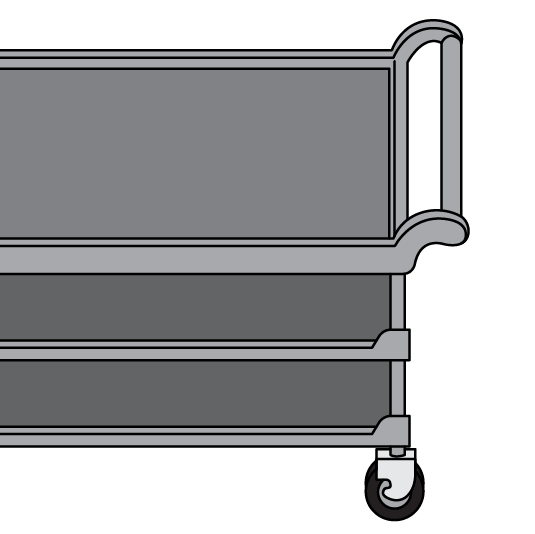}
  \includegraphics[width=0.1\textwidth]{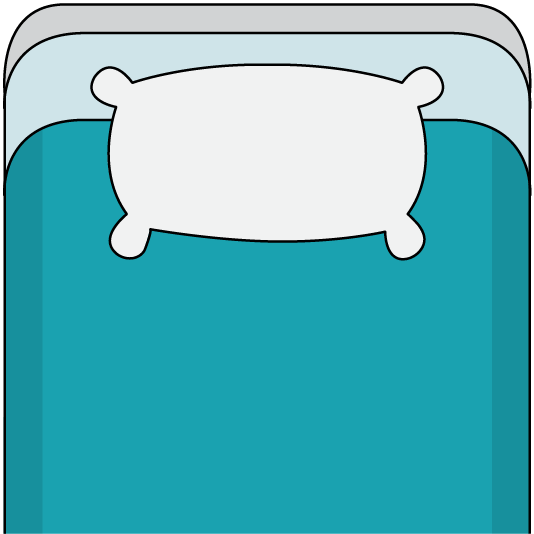}
  \includegraphics[width=0.1\textwidth]{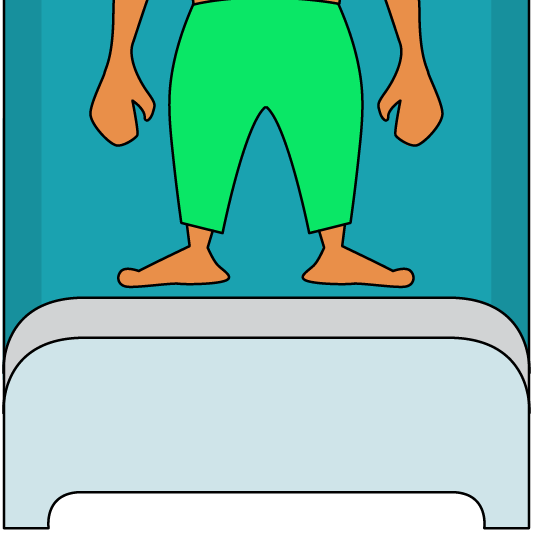}
  
  \caption{PDDL stations.}
  \label{fig:station_imgs}
\end{figure}

\textbf{Stations} are a class of immovable objects in the environment. Players can move to unoccupied cells adjacent to stations, as mentioned earlier, and items can be placed on stations. See available station assets in Figure \ref{fig:station_imgs}.

\begin{figure}[h]
  \centering
  \includegraphics[width=0.1\textwidth]{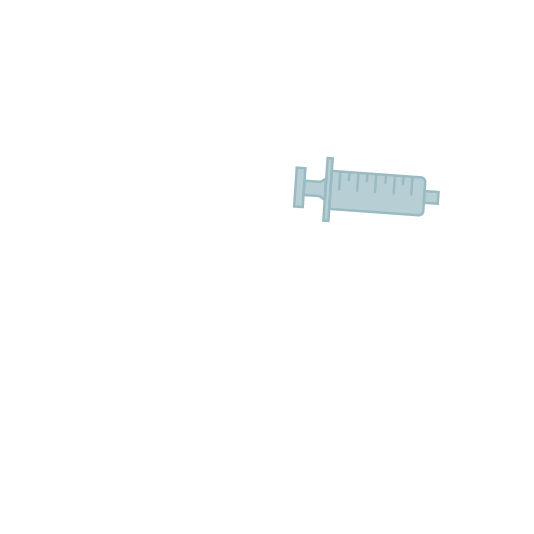}
  \includegraphics[width=0.1\textwidth]{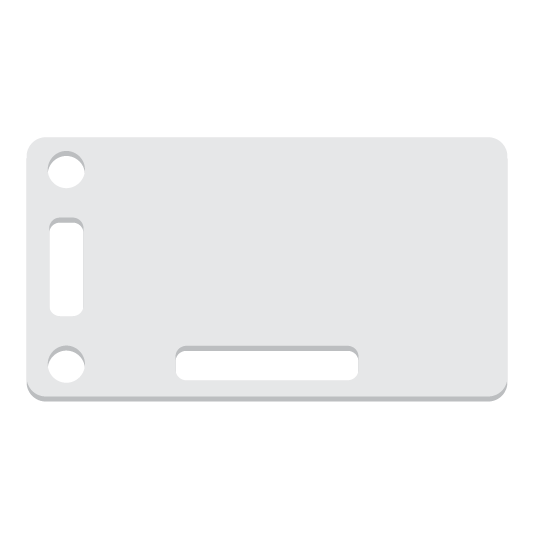}
  \includegraphics[width=0.1\textwidth]{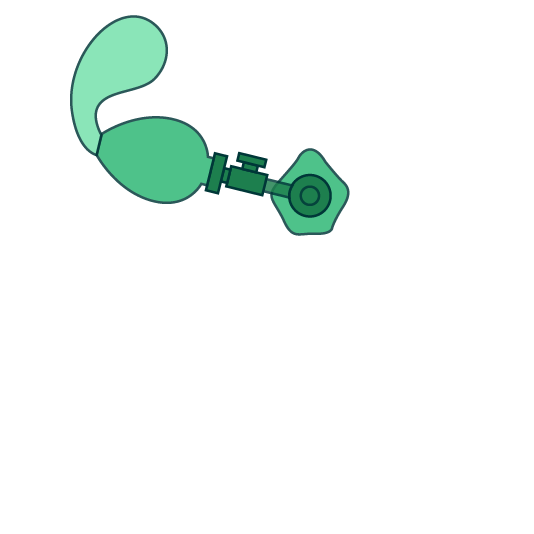}
  \includegraphics[width=0.1\textwidth]{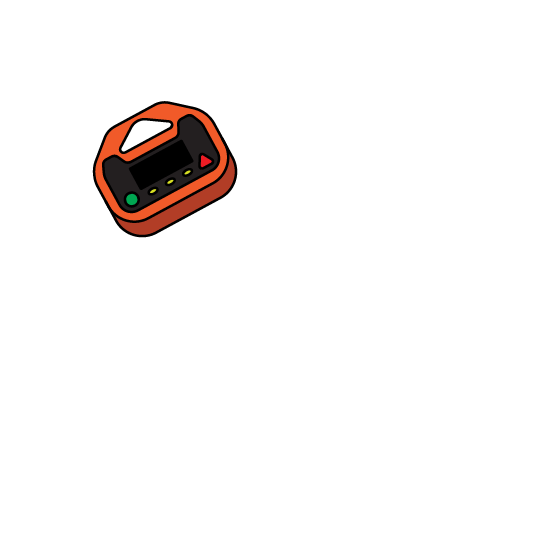}
  \hspace{0.5cm}
  \includegraphics[width=0.1\textwidth]{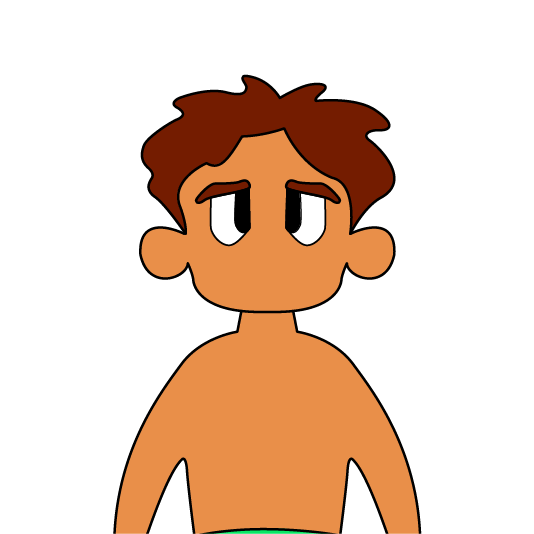}
  \includegraphics[width=0.1\textwidth]{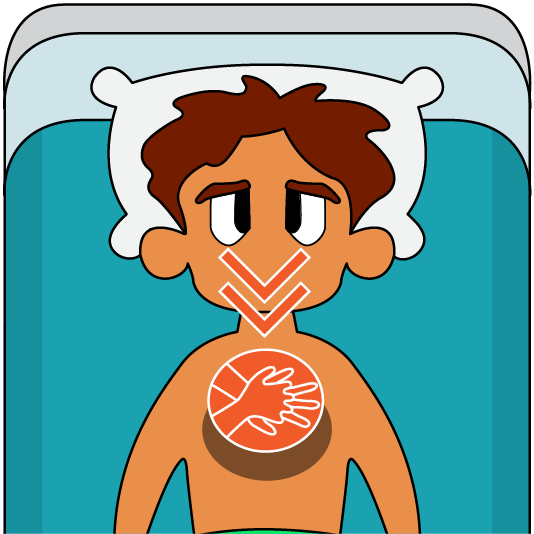}
  \includegraphics[width=0.1\textwidth]{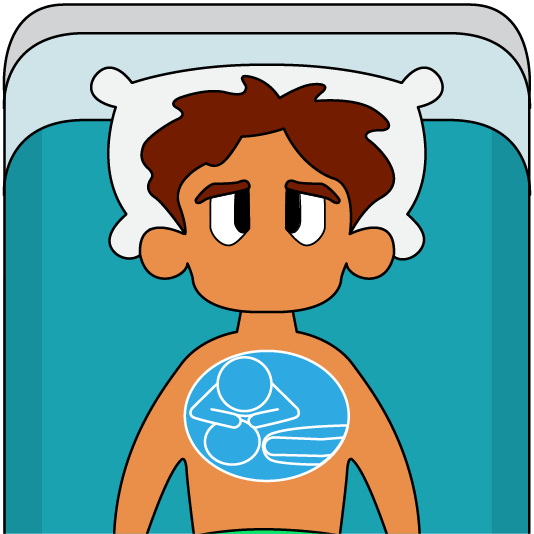}
  \includegraphics[width=0.1\textwidth]{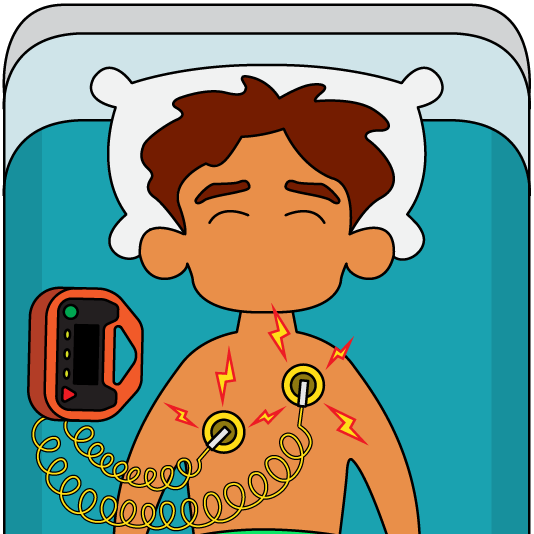}
  \includegraphics[width=0.1\textwidth]{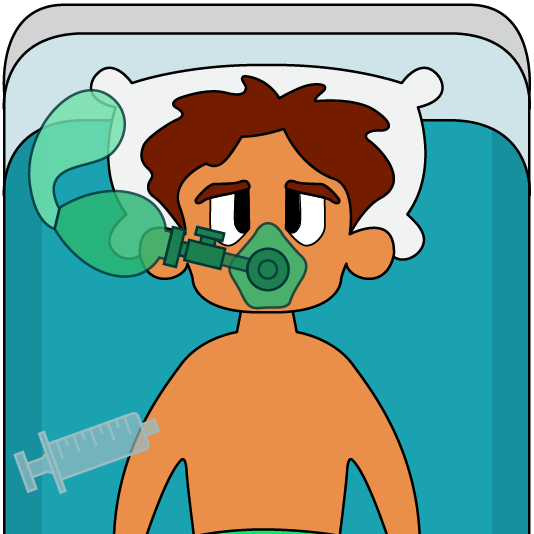}
  \caption{Item assets. \textit{Left}: Items used for treatment. From left to right: syringe, CPR board, breathing pump, AED device. \textit{Right}: Patient in various states. From left to right: default, after chest compressions, after rescue breaths, after AED shock, after given medicine}
  \label{fig:item_imgs}
\end{figure}

\textbf{Items} are interactable objects in the environment. Apart from the patient, the items in MARLHospital are tools necessary for the HCWs to complete tasks in treating the patient. They can be picked up and placed down by Players on stations. Items can be stacked on top of one another, though the player can only pick up the item at the top of a stack. To perform treatment actions, certain items need to be stacked on the patient based on that treatment action: a CPR board for \texttt{compress\_chest}, a breathing pump for \texttt{give\_rescue\_breaths}, an AED device for \texttt{giving\_shock}, and a syringe for \texttt{giving\_medicine}.

There are two items with special properties in the environment. The first is the CPR board, which, unlike other items, can also be stacked under the patient. The second is the patient itself. Unlike other items, the patient cannot be moved from the patient bed station, and it is the only possible predicate of treatment actions. The patient is rendered with different states based on the most recent task completed on them. See Figure \ref{fig:item_imgs} for item and patient assets.

\paragraph{Observation Space Size.}
The observation space consists of a total of 174 boolean values. Each of the three agents receives an individual observation vector of size 58. These 58 boolean features are structured as follows:

\begin{itemize}
    \item Patient State (4 booleans): Indicates whether the patient has been chest compressed, rescue breathed, treated, or shocked.
    
    \item \textbf{Agent Locations (18 booleans):}Encodes the location of each of the three agents, where each agent can be at one of six discrete stations: \texttt{hospital\_cart\_right1}, \texttt{table1}, \texttt{hospital\_cart1}, \texttt{hospital\_cart\_left1}, \texttt{patient\_legs1}, or \texttt{patient\_bed\_station1}.
    
    \item \textbf{Held Items (9 booleans):} Indicates which item (if any) is held by each of the three agents. Possible items include: \texttt{pump1}, \texttt{cpr\_board1}, and \texttt{patient1}.
    
    \item \textbf{Skill Levels (18 booleans):} Represents each agent’s skill level (unskilled, beginner, expert) for two skills: chest compression and rescue breathing.
    
    \item \textbf{Available Actions (9 booleans):} Encodes which actions are currently available to the agent, such as treating the patient, moving to any of the six stations, moving an item, or stacking an item under another.
\end{itemize}

\section{MARLHospital Environment}
\label{app:MARLHospital}

\FloatBarrier
\begin{figure}[!htb]
  \centering
  % \columnwidth is exactly one column's width in two‐col mode
  \includegraphics[width=0.9\columnwidth]{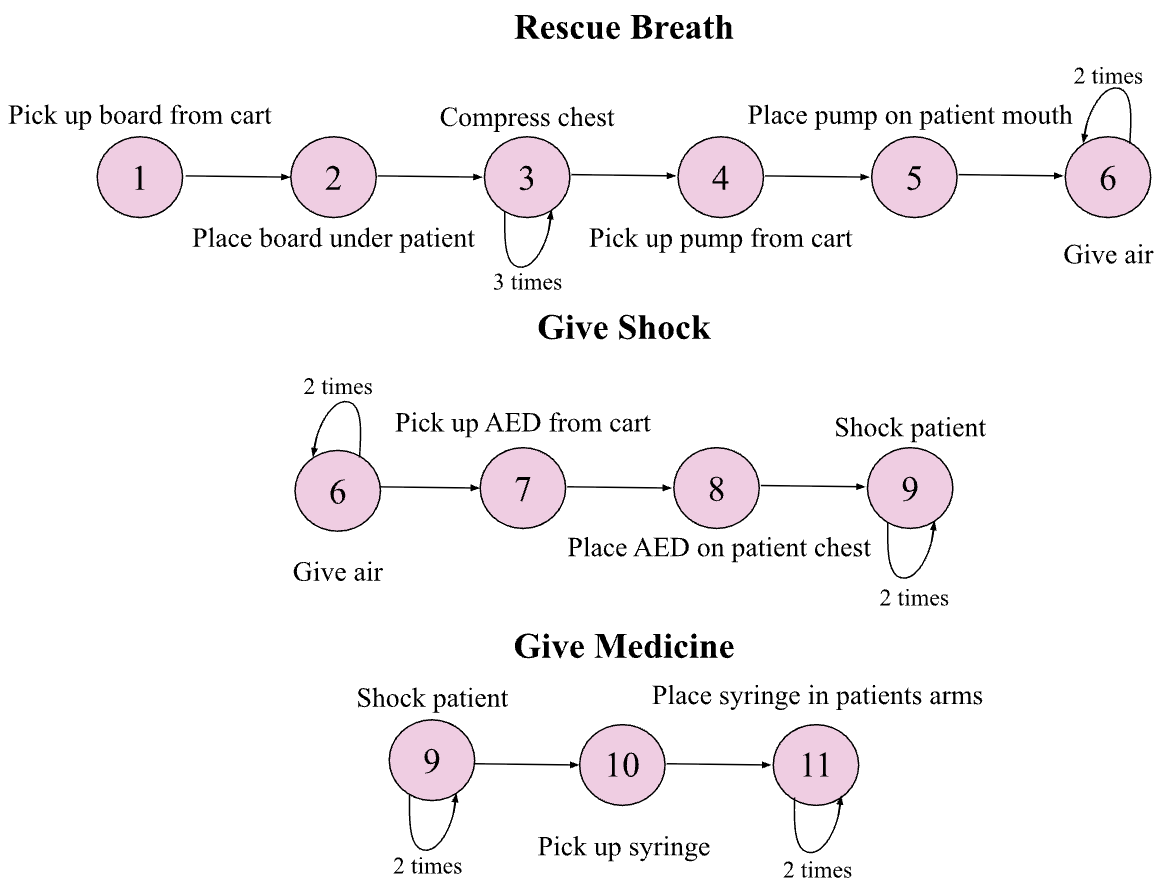}
  \caption{MARLHospital environment: Breakdown of the PDDL environment wrapped by the MARL environment with 3 HCW agents and a patient.}
  \label{fig:task-difficulty}
\end{figure}
\FloatBarrier

% \section{MARLHospital Environment}
% \label{app: MARLHospital}

% \begin{figure}[h]
%     \centering
%     \includegraphics[width=0.7\textwidth]{images/task_difficulty2.png} 
%     \caption{MARLHospital environment: Breakdown of the PDDL environment wrapped by the MARL environment with 3 HCW agents and a patient.}
%     \label{fig:main figure draft}
% \end{figure}

\begin{table*}[t]
\centering
\begin{ThreePartTable}
\caption{Comparison between MARLHospital and other simulators}
\label{tab:algorithms_properties}
\small
\setlength{\tabcolsep}{4pt}
\begin{tabularx}{\textwidth}{@{}l*{7}{Y}@{}}
\toprule
& Multi-Agent & MARL & Hospital Setting & Skill Specialization & Variable Skills & Energy Levels & Fairness Support \\
\midrule
Overcooked-AI \citep{carroll_utility_2020}                        & \cmark & \xmark & \xmark & \xmark          & \xmark & \xmark & \xmark \\
Agent Hospital \citep{li_agent_2024}                              & \cmark & \xmark & \cmark & \cmark\tnote{1} & \xmark & \xmark & \xmark \\
MARL Robotic Surgery \citep{scheikl_cooperative_2021-1}           & \cmark & \cmark & \cmark\tnote{2} & \cmark & \xmark & \xmark & \xmark \\
MARL for Nurse Rostering \citep{zhang_multi-agent_2024}           & \cmark & \cmark & \cmark\tnote{3} & \xmark & \xmark & \xmark & \xmark \\
Cuisine World \citep{gong_mindagent_2023}                         & \cmark & \xmark & \xmark & \xmark          & \xmark & \xmark & \xmark \\
VirtualHome \citep{puig_virtualhome_2018}                         & \cmark & \xmark & \xmark & \xmark          & \xmark & \xmark & \xmark \\
SMACv2 \citep{ellis_smacv2_2023}                                  & \cmark & \cmark & \xmark & \cmark          & \xmark & \xmark & \xmark \\
Pommerman \citep{resnick_pommerman_2018}                          & \cmark & \cmark & \xmark & \xmark          & \xmark & \xmark & \xmark \\
Melting Pot 2.0 \citep{agapiou_melting_2023}                      & \cmark & \cmark & \xmark & \cmark          & \xmark & \xmark & \xmark \\
Robotouille \citep{gonzalez-pumariega_robotouille_2025}           & \cmark & \xmark & \xmark & \xmark          & \xmark & \cmark & \xmark \\
HMARL for Medical Allocation \citep{hao_hierarchical_2023}         & \cmark & \cmark & \xmark & \xmark          & \xmark & \xmark & \xmark \\
MA Hospital Infection Sim \citep{esposito_multi-agent_2020}       & \cmark & \xmark & \cmark & \cmark          & \xmark & \xmark & \xmark \\
\midrule
\textit{MARLHospital}                                             & \cmark & \cmark & \cmark & \cmark          & \cmark & \cmark & \cmark \\
\bottomrule
\end{tabularx}
\begin{TableNotes}[flushleft]
\footnotesize
\item[1] Patients and doctors present, but only doctors are trained; no specialization among HCWs.
\item[2] A procedure occurs in a hospital context, but the environment does not map a physical hospital space.
\item[3] Combinatorial optimization for nurse rosters; does not cover patient treatment.
\end{TableNotes}
\end{ThreePartTable}
\end{table*}

\section{Selected Hyperparameters}
\label{app:hyper}

For hyperparameter tuning in MARLHospital, we primarily conducted a grid search and did a sweep across several hyperparameters for all the algorithms, ensuring consistency with established MARL benchmarks. The hyperparameter selection process follows a structured approach that is compatible with prior algorithms' work.

The hyperparameters for each algorithm, including IQL, %IPPO, 
MAPPO, VDN, and QMIX, are outlined in Tables \ref{tab:hyper_iql} to \ref{tab:hyper_qmix}. Key parameters include the hidden dimension size, learning rate, reward standardization, network type (e.g., GRU or fully connected networks), target update frequency, entropy coefficient (for policy gradient methods), and the number of steps used in bootstrapping (n-step returns).

For off-policy algorithms such as IQL, VDN, and QMIX, an experience replay buffer is employed to decorrelate samples and stabilize learning, following standard practices \cite{mnih_human-level_2015}. On-policy algorithms like %IPPO and 
MAPPO utilises parallel synchronous workers to mitigate sample correlation and improve stability during training.

The hyperparameter configurations for the IQL algorithm, with parameter sharing for MARLHospital is shown in Table \ref{tab:hyper_iql}.

\begin{table}[H]
    \centering
    \caption{Hyperparameters for IQL with parameter sharing.}
    \begin{tabular}{l c}
    \toprule
     {} & MARLHospital 
     \\
     \midrule
        hidden dimension & 64 \\
        learning rate  & 0.0005 \\
        reward standardisation & False \\
        network type & GRU \\
        evaluation epsilon & 0.0 \\
        target update & 20 (hard) \\
     \bottomrule 
    \end{tabular}
    \label{tab:hyper_iql}
\end{table}

% The hyperparameter configurations for the IPPO algorithm, with parameter sharing for MARLHospital is shown in Table \ref{tab:hyper_ippo}.

% \begin{table}[H]
%     \centering
%     \caption{Hyperparameters for IPPO with parameter sharing.}
%     \begin{tabular}{l c}
%     \toprule
%      {} & MARLHospital \\
%      \midrule
%         hidden dimension & 256 \\
%         learning rate  & 0.004 \\
%         reward standardisation & True \\
%         network type & GRU \\
%         entropy coefficient & 0.02 \\
%         target update & 0.08 (soft) \\
%         batch size & 6144 \\
%         buffer size & 150000 \\
%         $clipping coef$ & $0.5$ \\
%      \bottomrule 
%     \end{tabular}
%     \label{tab:hyper_ippo}
% \end{table}

“The ‘target update 0.01 (soft)’ in our 
% IPPO and 
MAPPO implementations is based on Polyak averaging, where the target critic network is updated slowly using 1`\%` new and 99 old weights for stability. Although we did not add it to the hyperparameter table, the PPO clipping coefficient is set to 0.2 and is used to constrain policy updates during training.

The hyperparameter configurations for the MAPPO algorithm, with parameter sharing for MARLHospital is shown in Table \ref{tab:hyper_mappo}.

\begin{table}
    \centering
    \caption{Hyperparameters for MAPPO with parameter sharing.}
    \begin{tabular}{l c}
    \toprule
     {} & MARLHospital \\
     \midrule
        hidden dimension & 64 \\
        learning rate  & 0.002 \\
        reward standardisation & True \\
        network type & GRU \\
        entropy coefficient & 0.01 \\
        target update & 0.05 (soft) \\
        clipping coef & $0.2$ \\
     \bottomrule 
    \end{tabular}
    \label{tab:hyper_mappo}
\end{table}

The hyperparameter configurations for the VDN algorithm, with parameter sharing for MARLHospital is shown in Table \ref{tab:hyper_vdn}.

\begin{table}
    \centering
    \caption{Hyperparameters for VDN with parameter sharing.}
    \begin{tabular}{l c}
    \toprule
     {} & MARLHospital \\
     \midrule
        hidden dimension & 64 \\
        learning rate  & 0.001 \\
        reward standardisation & False \\
        network type & GRU \\
        evaluation epsilon & 0.0 \\
        target update & 25 (hard) \\
     \bottomrule 
    \end{tabular}
    \label{tab:hyper_vdn}
\end{table}

The hyperparameter configurations for the QMIX algorithm, with parameter sharing for MARLHospital, are shown in Table \ref{tab:hyper_qmix}.

\begin{table}
    \centering
    \caption{Hyperparameters for QMIX with parameter sharing.}
    \begin{tabular}{l c}
    \toprule
     {} & MARLHospital \\
     \midrule
        hidden dimension & 64 \\
        learning rate  & 0.001 \\
        reward standardisation & False \\
        network type & GRU \\
        evaluation epsilon & 0.1 \\
        target update & 25 (hard) \\
     \bottomrule 
    \end{tabular}
    \label{tab:hyper_qmix}
\end{table}

\begin{figure}[h]
    \centering
    \framebox{\includegraphics[width=0.35\textwidth]{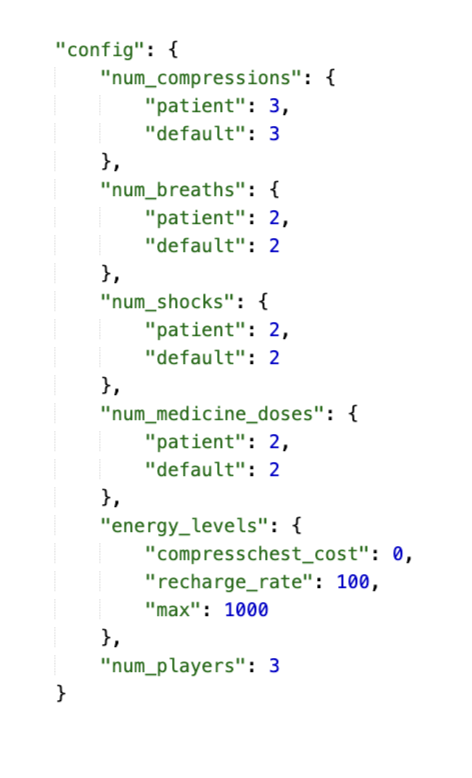}}
    \caption{JSON to configure the necessary steps to treat the patient}
    \label{fig:config}
\end{figure}
\begin{center}
    
\end{center}

% \FloatBarrier
% \begin{figure}[!htb]
%   \centering
%   \framebox{%
%     \includegraphics[width=0.9\columnwidth]{images/givemedicine_equal.png}%
%   }
%   \caption{JSON to configure the HCWs skill information for equal skill.}
%   \label{fig:equal}
% \end{figure}
% \FloatBarrier

% \FloatBarrier
% \begin{figure}[!htb]
%   \centering
%   \framebox{%
%     \includegraphics[width=0.9\columnwidth]{images/givemedicine_spec.png}%
%   }
%   \caption{JSON to configure the HCWs skill information for specialized roles.}
%   \label{fig:specialized}
% \end{figure}
% \FloatBarrier

\clearpage
\onecolumn

\begin{figure}[htbp]
    \centering
    \framebox{\includegraphics[width=0.8\textwidth]{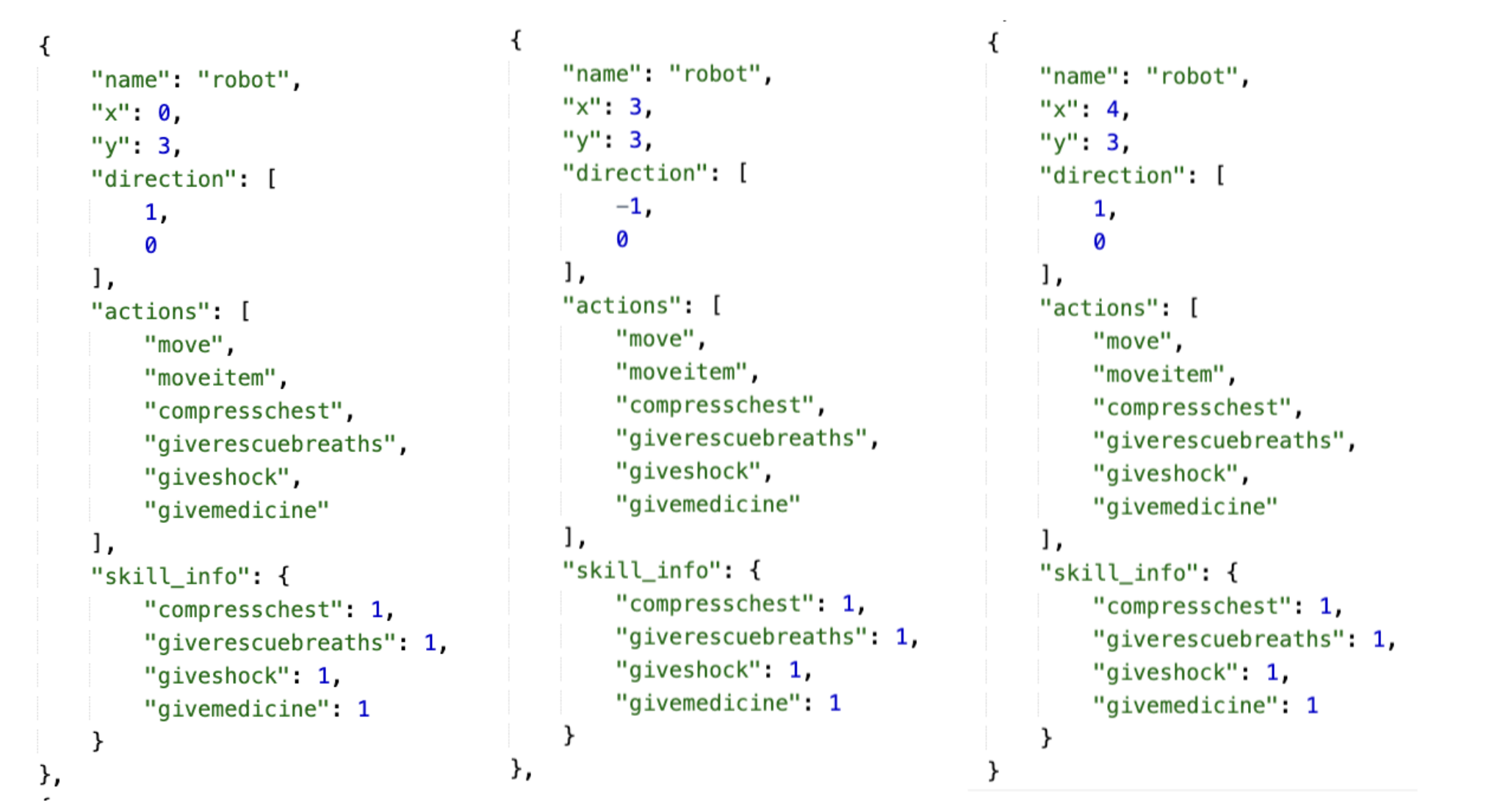}}
    \caption{JSON to configure the HCWs skill information for equal skill.}
    \label{fig:equal}
\end{figure}

\begin{figure}[htbp]
    \centering
    \framebox{\includegraphics[width=0.7\textwidth]{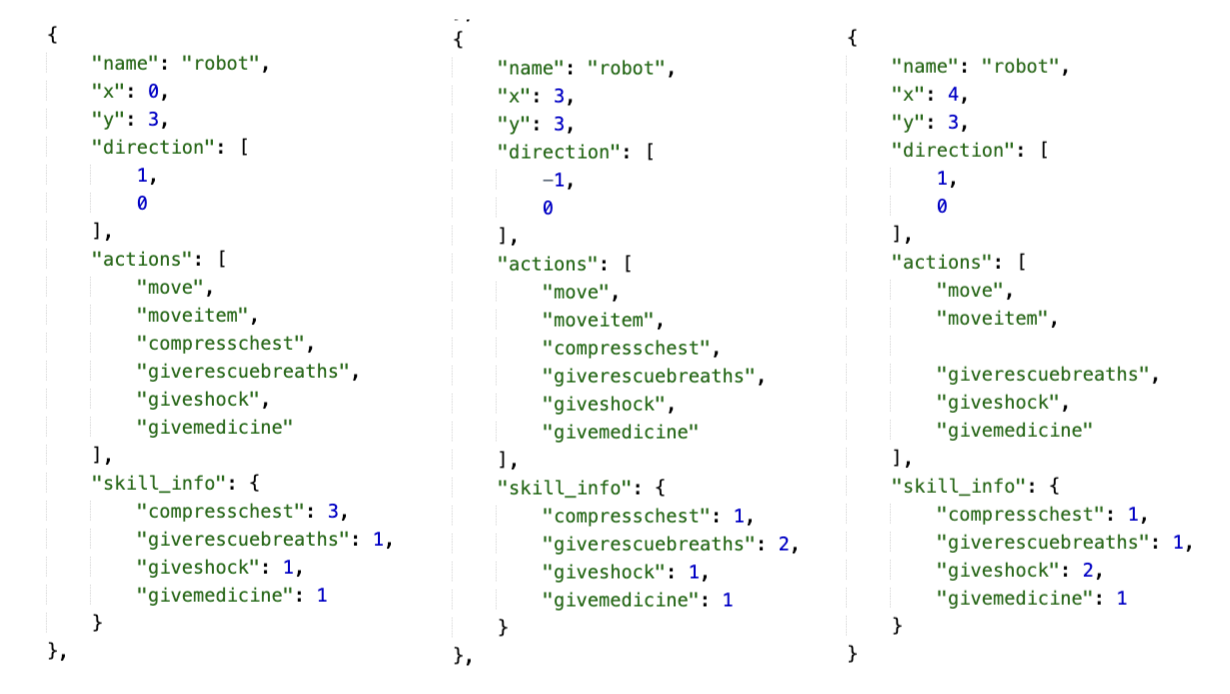}}
    \caption{JSON to configure the HCWs skill information for specialized roles.}
    \label{fig:specialized}
\end{figure}

\begin{figure}[htbp]
    \centering
    \framebox{\includegraphics[width=0.9\textwidth]{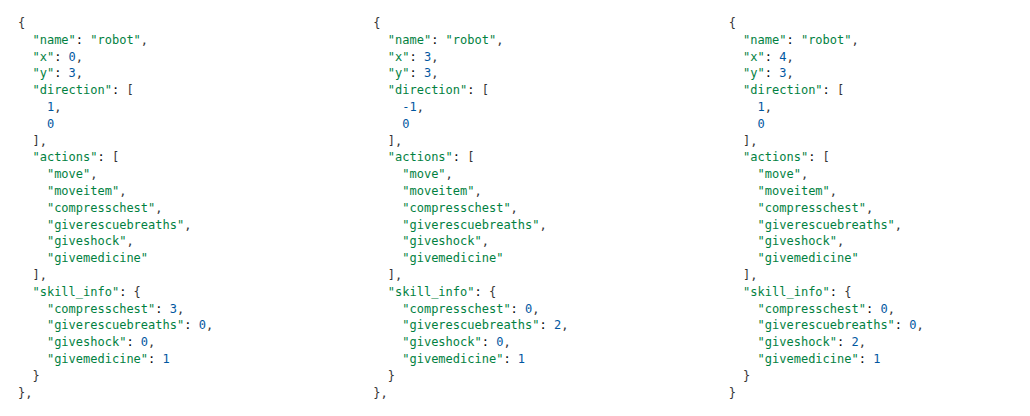}}
    \caption{JSON to configure the HCWs skill information for required cooperation}
    \label{fig:forcedcoop}
\end{figure}

\clearpage
\twocolumn

% \begin{figure}[h]
%     \centering
%     \framebox{\includegraphics[width=1\textwidth]{images/givemedicine_equal.png}}
%     \caption{JSON to configure the HCWs skill information for equal skill}
%     \label{fig:equal}
% \end{figure}
% \begin{center}
    
% \end{center}

% \begin{figure}[h]
%     \centering
%     \framebox{\includegraphics[width=1\textwidth]{images/givemedicine_spec.png}}
%     \caption{JSON to configure the HCWs skill information for specialized roles}
% \end{figure}
% \begin{center}
    
% \end{center}

\clearpage

\end{document}